\documentclass[pra,times,amsmath,amssymb,twocolumn,showpacs,superscriptaddress]{revtex4}
\usepackage[usenames]{color}
\usepackage{amssymb,hyperref}
\usepackage{grffile}
\usepackage[pdftex]{graphicx}
\usepackage{amsmath, amstext, amssymb, amsfonts, amsxtra}
\usepackage{textcomp}
\usepackage{xspace}
\usepackage{icomma}
\usepackage[short]{datetime}
\usepackage[normalem]{ulem}
\usepackage{pdfpages}

\def \su{{\uparrow}}
\def \sd{{\downarrow}}
\def \sud{{\uparrow\downarrow}}
\def \nb{\bar{n}}
\def \up{{\uparrow}}
\def \down{{\downarrow}}

\def \ell{{d}}

\newcommand{\aver}[1]{\ensuremath{\langle #1 \rangle}\xspace}%

\newcommand{\gzeno}{{\lambda}_{n,\text{Zeno}}}   
\newcommand{\gad}{{\gamma}_{n,\text{ad}}}

\def \r{ {\bf{r}} }

\newcommand{\bonn}{University of Bonn, HISKP, Nussallee 14-16, 53115 Bonn, Germany}
\newcommand{\ubc}{Department of Physics and Astronomy, University of British Columbia, Vancouver V6T 1Z1, Canada.} 

\newcommand{\sutd}{Singapore University of Technology and Design, 20 Dover Drive, 138682 Singapore.}
\newcommand{\pulse}{Pulse Energy, 600-576 Seymour Street, Vancouver V6B 3K1, Canada.}

\begin{document}

\title{Dissipative quantum dynamics of fermions in optical lattices: a slave-spin approach}

\author{Jean-S\'ebastien Bernier}
\affiliation{\ubc}
\affiliation{\pulse}
\author{Dario Poletti}
\affiliation{\sutd} 
\author{Corinna Kollath}
\affiliation{\bonn}

\begin{abstract}
We investigate the influence of a Markovian environment on the dynamics of interacting spinful
fermionic atoms in a lattice. In order to explore the physical phenomena occurring at short times, 
we develop a method based on a slave-spin representation of fermions which is 
amenable to the investigation of the dynamics of dissipative systems. 
We apply this approach to two different dissipative couplings which can occur in 
current experiments: a coupling via the local density
and a coupling via the local double occupancy. We complement our study based on this  
novel method with results obtained using the adiabatic elimination technique and with 
an exact study of a two-site model. We uncover that the decoherence is slowed down by increasing either 
the interaction strength or the dissipative coupling (the Zeno effect). We also find, for 
the coupling to the local double occupancy, that the final steady state can sustain 
single-particle coherence.  
\end{abstract}


%05.40.Ca Noise
%03.75.Ss Degenerate Fermi gases 
%03.65.Yz Decoherence; open systems; quantum statistical methods
%05.30.Fk Fermion systems and electron gas
%67.85.-d Ultracold gases, trapped gases
%67.85.De, Dynamical properties of BEC 
%67.85.Hj  Bose-Einstein condensates in optical potentials 

\maketitle

%%%%%%%%%%%%%%%%%%%%%%
% introduction
%%%%%%%%%%%%%%%%%%%%%%

\section{Introduction}

The behavior of physical systems is typically influenced by the surrounding environment. However, identifying the 
correct environmental mechanisms and their corresponding effects is quite often a difficult task. 
For example, the relevant environment can be as simple as external noises, like magnetic field fluctuations, but can also 
be system specific couplings such as electrons coupled to phononic modes in solids or cold atomic gases subjected 
to fluorescence scattering in near resonant optical lattices~\cite{footnote1}.
%\footnote{Fluorescence scattering is the absorption of 
%a (lattice) photon by an atom which is then followed by a spontaneous emission.}. 
In fact, a large corpus of theoretical frameworks has been developed to describe
the interactions between a system and its environment in various limits. 
This vast inventory ranges from the description of electron-phonon
couplings using baths of harmonic oscillators, to the Markovian description of memory-less environments, 
a prevalent approach in the field of quantum optics.
In recent years, the renewed interest for the physics of environmentally coupled systems has been fueled
by the realization that environments can perturb non-trivially the central system by inducing, for example,
superconductivity in solid state systems~\cite{FaustiCavalleri2011} or Zeno- and anti-Zeno effects in quantum optical systems~\cite{StreedPritchard2006}. 
Furthermore, state preparation by environmental coupling, which optical pumping is one of the most common
implementations, has recently been generalized and promises to be a great avenue to realize intriguing states 
of matter~\cite{SyassenDuerr2008, BarreiroBlatt2010, MuellerZoller2012, CaballarWatanabe2014, BernierKollath2013, Daley2014,BrennenWilliams2003}. 
The interest for these systems is compounded by the
prediction of surprising effects such as the impeding of decoherence by dissipative coupling in bosonic gases under dephasing or atom losses~\cite{PichlerZoller2010,PolettiKollath2012,ShchesnovishKonotop2010, BrazhnyiOtt2009, WitthautWimberger2011, Daley2014}, 
the enhancement or appearance of distinct transport properties under the influence of globally acting dissipation~(see e.g.~\cite{MendozaArenasClark2013,KesslerMarquardt2012} and references therein), the presence of scaling regimes in the dissipative quantum dynamics~\cite{PolettiKollath2012,PolettiKollath2013, CaiBarthel2013, TomadinZoller2011, WitthautWimberger2011}, and the occurrence of glass-like dynamics in dissipatively coupled many-body systems~\cite{PolettiKollath2013,OlmosGarrahan2012,LesanovskyGarrahan2013}.
Despite these impressive advances, the interplay between the environmental coupling and the interacting many-body dynamics, 
and by extension the induced effects, are still not fully understood.

The development of a better comprehension for these systems, in particular for fermionic ones, has historically
been hindered by a lack of proper theoretical methods to explore their rich physical structure. We develop here
a novel method, based on a slave-spin representation of fermions~\cite{deMediciBiermann2005}, to uncover the physics of many-body systems
coupled to Markovian environments. This newly developed approach proves to be extremely useful to investigate the short-time
dynamics of a broad class of strongly interacting fermionic systems. We show its applicability to two different kinds of
dissipative processes: environmental effects described by couplings to either the system density or double-occupancy. 
The coupling to the density of fermions is one of possible dissipative effects that can occur in ultracold 
fermionic gases in red-detuned optical lattices~\cite{GerbierCastin2010,PichlerZoller2010} by fluorescence scattering.
 In order to engineer the coupling to the double occupancy, an interaction energy shift can be added~\cite{BrennenWilliams2003},
 such that only doubly occupied sites are sensitive to a certain driven transition.
Thus, our findings have wide implications for the stability of complex quantum states in realizable experiments. 
We also supplement the developed approach with comparisons to exact diagonalization and adiabatic 
elimination methods~\cite{GardinerZollerBook, CarmichaelBook,BreuerPetruccione2002} applied to
the same dissipative many-body systems.

The rest of the article is structured as follows: in section \ref{sec:model}, we present the model and the 
considered dissipative couplings with their corresponding asymptotic long-time steady states. In section \ref{sec:methods}, 
we introduce the different theoretical approaches. We describe in \ref{sec:ed} the exact diagonalization approach, in 
\ref{sec:slave} the newly developed slave-particle approach, and in \ref{sec:adiabatic} the adiabatic elimination method. 
In section \ref{sec:dissipativeevolution}, we present our results for the coupling to the two different environments: in \ref{sec:density}, 
the coupling to a local dephasing noise is investigated, whereas in \ref{sec:doublons} we consider the local coupling 
to doubly occupied sites. The various arising effects, the Zeno effect, the slowing down of decoherence 
and its oscillating behavior in the presence of strong interaction, are discussed within this section. Finally, 
we conclude in section \ref{sec:conclusions}.

%%%%%%%%%%%%%%%%%%%%%%%
% modeling of the system
%%%%%%%%%%%%%%%%%%%%%%
\section{Description of the system and its steady state}\label{sec:model}
\subsection{Model}
We investigate the Hamiltonian evolution of interacting fermions in a lattice potential described 
by the $d$-dimensional Hubbard model 
\begin{eqnarray} 
\label{eq:Hubb_Ham}
H \!\! &=& \!\! -J\!\!\!\sum_{{\bf r}, {\bf r}'({\bf r}),\sigma} 
      \!\left( c_{{\bf r},\sigma}^\dagger  c^{\phantom{\dagger}}_{{\bf r}',\sigma}+\mbox{h.c.}\right) \\
&& \,\,\,\,\,\,\,\,\,\,\,\,\,\,\,\,\,\,\,\,
+ \frac{U}{2}~\sum_{\bf r} (n_{{\bf r},\uparrow}+n_{{\bf r},\downarrow}-1)^2 \nonumber
\end{eqnarray}
where $ c_{\r,\sigma}^\dagger$ is the creation operator for a fermion with 
spin $\sigma = \{\uparrow\!,\downarrow \}$ and site index $\r$, ${n}_{\r,\sigma} =  c_{\r,\sigma}^\dagger c^{\phantom{\dagger}}_{\r,\sigma}$ 
is the density operator. Here ${\bf r}'({\bf r})$ denotes all nearest-neighbors of site ${\bf r}$.
The first term in Eq.~(\ref{eq:Hubb_Ham}) corresponds to the kinetic energy of the fermions and $J>0$ is the hopping coefficient. 
The second term characterizes the repulsive interaction with strength $U>0$. This model is often used 
to describe fermionic atoms confined to optical lattice potentials~\cite{KoehlEsslinger2005, BlochZwerger2008} 
or electrons in solids~\cite{AshcroftMermin}. 

In addition to the unitary dynamics, we consider the presence of different couplings to the environment causing 
a dissipative evolution. For Markovian processes, the evolution of the density matrix, $\rho$, describing the fermions 
follows a master equation 
\begin{equation}
\frac{d}{dt} \rho(t)=-\frac {\rm i} \hbar \big[H,\, \rho(t)\big]
+ \mathcal{D}\left(\rho\left(t\right)\right)\,. \label{eq:master}     
\end{equation}
The first term on the right hand side of Eq.(\ref{eq:master}) corresponds to the unitary evolution induced by 
the Hamiltonian. The second term models the dissipative coupling. The effects of the considered environments are described by dissipators within the Lindblad formalism, i.e.  
\begin{eqnarray}
\label{eq:dissipator}
\mathcal{D}(\rho) &=& \gamma~\sum_{{\bf r},\sigma} \left( j_{{\bf r},\sigma}\rho~j^\dagger_{{\bf r},\sigma} 
- \frac{1}{2} j^\dagger_{{\bf r},\sigma} j_{{\bf r},\sigma} \rho - \frac{1}{2} \rho~j^\dagger_{{\bf r},\sigma}j_{{\bf r},\sigma} \right) \nonumber \\
\end{eqnarray}
where $j_{{\bf r},\sigma}$ denotes the quantum jump operators. 

We consider here two types of dissipative environment. These environments, chosen for the interesting dynamical effects
they induce, are described within the Lindblad formalism using the following
quantum jump operators:
\begin{itemize}
\item[(i)] the local density 
\begin{equation}
\label{eq:dissn}
j_{{\bf r},\sigma}=n_{{\bf r},\sigma};
\end{equation}
\item[(ii)] the local double occupancy density 
\begin{equation}
\label{eq:dissd}
j_{{\bf r},\sigma}=\frac{1}{\sqrt{2}} n_{{\bf r},\uparrow}n_{{\bf r},\downarrow}.
\end{equation}
\end{itemize}
In the analysis that follows, we typically consider the situation where the 
system is initially prepared in the ground state of the Fermi-Hubbard model, and then the coupling to the environment is 
switched on. Our study focuses on the time evolution of the system which, as we will present in details, 
strongly depends on the interplay of the unitary and dissipative dynamics.  

\subsection{Structure of the decoherence free subspaces and steady states}
\label{sec:steadystates}
The two different dissipative couplings exemplify well two distinct situations:
\begin{itemize}
\item[(i)] the local density coupling has a unique steady state towards which {\it any} initial state evolves; 
\item[(ii)] the local double occupancy coupling has an entire subspace of steady states and the asymptotic long time 
state depends on the initial condition. 
\end{itemize}

For case (i) the decoherence free subspace of the dissipator, defined by $\mathcal{D}(\rho)=0$, is given by all density matrices 
which are diagonal within the Fock basis. For these density matrices the first term in the dissipator Eq.~(\ref{eq:dissipator}) 
is exactly canceled by the last two terms. This implies that without the Hamiltonian evolution, i.e.~$H=0$, the entire 
set of diagonal density matrices are steady states ($\dot{\rho}=0$). While the interaction part of the Hamiltonian leaves these 
states untouched, the competing kinetic term lifts this degeneracy. As a consequence, only the infinite temperature state, i.e. 
the state proportional to unity in the Fock basis, is invariant under the combined action of the dissipator and of the 
Hamiltonian. As we will later use the representation of this steady state as a reduced single site density matrix $\tilde{\rho}$ 
(where one traces out all sites except one), we provide here its definition in the single site Fock basis 
\begin{eqnarray}
\tilde{\rho}(t)&=&\sum_{m=0,\uparrow, \downarrow,\uparrow \downarrow}  \tilde{\rho}_{m}(t) |m\rangle\langle m|\\
 \tilde{\rho}_{0}(t=\infty)%=p_{\su}(0)p_{\sd}(0)
&=&(1-\nb_{\su})(1-\nb_{\sd}) \label{eq:ss1}  \\  
 \tilde{\rho}_{\su}(t=\infty)%=p_{\su}(1)p_{\sd}(0)
&=&\nb_{\su}(1-\nb_{\sd}) \label{eq:ss2}  \\  
 \tilde{\rho}_{\sd}(t=\infty)%=p_{\su}(0)p_{\sd}(1)
&=&(1-\nb_{\su})\nb_{\sd} \label{eq:ss3}  \\  
 \tilde{\rho}_{\sud}(t=\infty)%=p_{\su}(0)p_{\sd}(0)
&=&\nb_{\su}\nb_{\sd}. \label{eq:ss4}  
\end{eqnarray}
where $\nb_ \sigma$ denotes the average density for the fermions with spin $\sigma$. It is important to note that we 
are considering here an infinite homogeneous system in any dimension. 
 
For case (ii), where the environment effectively couples to the local double occupancy, the situation is very different. 
Here the decoherence free subspace is much larger. It contains the diagonal matrices in Fock space, but additionally, 
states which have no double occupancies, but still coherence. Since the interaction does not affect this subclass of states, 
all states belonging to this subspace and that are additionally eigenstates of the kinetic term of the Hamiltonian are steady states. 
The infinite temperature state is one of these steady states. However, the actual steady state reached depends on the overlap of the 
initial wavefunction with the possible steady states and can thus vary. We will discuss this 
situation in more detail below using the example of two fermions on two sites where almost analytical solutions 
are found. The occurrence of this entire subspace of steady states induces a very interesting long time dynamics. 
However, in this work we put the emphasis on the short time dynamics which is more relevant to current experiments, and 
we postpone further discussions on the long time dynamics to future works. 

\section{Methods}\label{sec:methods}
\subsection{Eigenvalues and eigenstates for the master equation}\label{sec:ed}
The evolution equation \eqref{eq:master} is a linear with respect to the density matrix. It can thus be rewritten in the 
matrix form
\begin{equation}
\partial_t\rho=M \rho. 
\end{equation}
Here $\rho$ is reordered as a vector and the elements of the matrix $M$ are chosen to be identified with the right-hand side 
of the master equation. Using this matrix form is advantageous as the time evolution of an initial 
density vector $\rho(t=0)$ is then related to the (complex) eigenvalues $\lambda_i$ and 
right eigenvectors $v_i$ of the non-Hermitian matrix $M$, i.e.
\begin{equation}
\rho(t)=\sum_i c_i e^{t\lambda_i} v_i.
\end{equation}
The weights $c_i$ are chosen such that the initial state can be represented as 
$\rho(t=0)=\sum_i c_i v_i$. One directly sees from this expression, that the real and imaginary parts of the eigenvalues 
lead to very different dynamics with time $t$. Whereas the imaginary part leads to an oscillatory behavior, a real negative 
contribution of the eigenvalues leads to an exponential decay. Due to these exponential decays, only the states having
null real parts will survive at long times and can potentially be steady states.  

\subsubsection{Application to a system of two fermions on two sites}
\label{sec:2sites2fermions}
In this subsection, we consider the simplified problem of two fermions, one with spin up and one with spin down, 
on two sites. This problem is, to a certain extent, treatable analytically and can already provide a lot of insight 
into the interplay of the hopping, interaction and dissipative terms. We will see in the discussion of the results 
in Sec.~\ref{sec:dissipativeevolution} that the main dynamical effects occurring on two sites are recovered 
in the extended system. For this system of two sites, we choose the Fock basis 
$(|\!\uparrow \downarrow, 0\rangle$, $|0, \uparrow \downarrow \rangle, |\!\uparrow, 
\downarrow \rangle |\!\downarrow, \uparrow \rangle)$. Within this basis the Hamiltonian is represented as 
\begin{equation}
H = \left( \begin{array}{cccc}
U & 0 & -J & J\\
0 & U & -J & J\\
-J & -J & 0 & 0\\
J & J & 0 & 0
\end{array} \right)
\end{equation}
and its ground state is given by
\begin{equation}
|GS\rangle = \frac{1}{A_1}\left( \begin{array}{c}
-4J\\
-4J\\
-U-\sqrt{16J^2+U^2}\\
 U+\sqrt{16J^2+U^2}
\end{array} \right)
\end{equation}
with $A_1= \frac{1}{\sqrt{2\left(U+\sqrt{16J^2+U^2}\right)^2+32J^2}}$.
The master equation can be understood as $16$ coupled differential equations and the density matrix as a vector with $16$ entries.
Even though this system is still quite complex, a lot of information can be extracted by diagonalizing the matrix $M$. 
This can typically not be fully performed analytically, but approximate analytical 
expressions are obtained for the dominating eigenvalues. In the following sections, we analyze the effect of the two 
aforementioned quantum jump operators.

\subsubsection{Solutions for case (i): dissipative coupling to the local density}
\label{sec:2sites2fermionsdensity}
In the case of two fermions on two sites where the effective dissipative coupling is to the local density,
the dissipator is described by  
\begin{equation}
\mathcal{D}_n(\rho) =  -\gamma \left( \begin{array}{cccc}
0 & 2\rho_{12} & \rho_{13} & \rho_{14} \\
2\rho_{21} & 0 & \rho_{23} & \rho_{24} \\
\rho_{31} & \rho_{32} & 0 & 2\rho_{34} \\
\rho_{41} & \rho_{42} & 2\rho_{43} & 0 
\end{array} \right) \label{eq:dissipatorn}
\end{equation} 
and the eigenvalues can be determined analytically. As we are mainly interested in analyzing time evolutions which begin from the ground
state of the Hamiltonian, we restrict our discussion to eigenmodes having an overlap with this ground state. The relevant 
eigenvalues and eigenstates are thus
\begin{eqnarray}
&& \lambda_{n,0} = 0, \;\;\;\; v_{n,0} = \frac{1}{4}\begin{pmatrix}1 & 0& 0& 0\\ 0& 1& 0& 0\\ 0& 0& 1& 0\\ 0& 0& 0& 1\end{pmatrix}, \label{eq:vn0} \\
&& \lambda_{n,1} = -2\gamma, \;\;\;\; v_{n,1} = \frac{1}{2}~\begin{pmatrix}0& -1& 0& 0\\ -1& 0& 0& 0\\ 0& 0& 0& 1\\ 0& 0& 1& 0\end{pmatrix}, \\
&& \lambda_{n,2} = -\gamma-\frac{1}{\sqrt{2}}\sqrt{-16 \left(\frac{J}{\hbar}\right)^2+\gamma^2-\left(\frac{U}{\hbar}\right)^2-A},\nonumber\\ \\
&& \lambda_{n,3} =-\gamma+\frac{1}{\sqrt{2}}\sqrt{-16 \left(\frac{J}{\hbar}\right)^2+\gamma^2-\left(\frac{U}{\hbar}\right)^2-A},\nonumber\\ \\
&& \lambda_{n,4} =-\gamma-\frac{1}{\sqrt{2}}\sqrt{-16 \left(\frac{J}{\hbar}\right)^2+\gamma^2-\left(\frac{U}{\hbar}\right)^2+A},\nonumber\\ \\
&& \lambda_{n,\text{Zeno}} = -\gamma+\frac{1}{\sqrt{2}}\sqrt{-16 \left(\frac{J}{\hbar}\right)^2+\gamma^2-\left(\frac{U}{\hbar}\right)^2+A}, \nonumber\\ \label{eq:d16} 
\end{eqnarray}
with $\quad A=\sqrt{-64\gamma^2\left(\frac{J}{\hbar}\right)^2+\left[16\left(\frac{J}{\hbar}\right)^2+\gamma^2+\left(\frac{U}{\hbar}\right)^2\right]^2}$. 
Only the eigenvectors corresponding to the first and second eigenvalues have an easy analytical form 
and have been presented above (in matrix form). The eigenvalue $\lambda_{n,0}$ is the only one to be null (for $J,~U,~\gamma~\ne 0$) which
means that the corresponding density matrix $v_{n,0}$, proportional to the identity, is the only 
asymptotic long-time state of the system. As the identity density matrix corresponds to the infinite temperature state or the totally mixed 
state of the system, the combined effect of the dissipation and of the Hamiltonian evolution is to heat up the system 
towards the infinite temperature state. In contrast, the state corresponding to eigenvalue $\lambda_{n,1}$ decays exponentially 
with a rate $2\gamma$. In turn, the form of the corresponding eigenstate $v_{n,1}$ implies that the coherence between the doubly occupied states  
$|\!\uparrow \downarrow, 0\rangle$ and $|0, \uparrow \downarrow \rangle$ and between 
the states $ |\!\uparrow, \downarrow \rangle$ and $|\!\downarrow, \uparrow \rangle)$ is fragile and decays 
exponentially. 

The eigenvalues $\lambda_{n,2}$ and $\lambda_{n,3}$ have an imaginary part for all $\frac{U}{J}> 0$ 
and $\frac{\hbar\gamma}{J} \ge 0$ which is, at large values of the interaction strength, proportional to $U$. 
While the imaginary parts cause oscillations, in the region where the overlap of the ground state with these states is
maximal, the characteristic damping time, $\tau_\text{damp} = \frac{1}{|\text{Re}(\lambda_{n,2})|}$, is much shorter than an oscillation 
period given by $T = \frac{2\pi}{|\text{Im}(\lambda_{n,2})|}$. As a consequence, in this regime the oscillations are overdamped. In contrast, the eigenvalues $\lambda_{n,4}$ and $\lambda_{n,Zeno}$ are real for all $\frac{U}{J}\ge 0$ and $\frac{\hbar\gamma}{J} \ge 0$. 
Therefore, the corresponding density matrices will only show a simple exponential decay. For small hopping elements $J \ll U,\;\hbar\gamma$, 
the eigenvalues can be well described by 
$$
\lambda_{n,4}= -2\gamma,
$$
and 
$$
\lambda_{n,\text{Zeno}}=-\frac{8\gamma J^2}{(\hbar\gamma)^2+U^2}.
$$
The density matrices corresponding to $\lambda_{n,4}$ will thus decay rapidly with a decay rate $2\gamma$. 
In contrast, the eigenvalue $\lambda_{n,\text{Zeno}}$ has only a contribution quadratic in the hopping element, and
consequently the corresponding density matrix decays very slowly. In particular, for large dissipative coupling the decay rate 
is approximately given by $-\frac{8J^2}{\hbar^2\gamma}$ which {\it decreases} with increasing dissipative coupling $\gamma$. 
This counterintuitive phenomenon is known as the Zeno effect and its origin will be discussed further in section \ref{sec:adiabatic}. 
For $U \gg \hbar\gamma$, the decay rate is approximated by $-\frac{8\gamma J^2}{U^2}$, this effect is known for bosonic systems as interaction impeding 
effect~\cite{PichlerZoller2010,PolettiKollath2012,PolettiKollath2013}. A similar slow time-scale arises 
in the presence of atom losses and has been named non-linear Zeno 
effect~\cite{ShchesnovishKonotop2010, BrazhnyiOtt2009, WitthautWimberger2011}.  

\subsubsection{Solutions for case (ii): dissipative coupling to the local double occupancy}
For the coupling to the local double occupancy, the dissipator can be written in the form 
\begin{equation}
\mathcal{D}_d (\rho)=  -\frac{\gamma}{2} \left( \begin{array}{cccc}
0 & 2\rho_{12} & \rho_{13} & \rho_{14} \\
2\rho_{21} & 0 & \rho_{23} & \rho_{24} \\
\rho_{31} & \rho_{32} & 0 & 0 \\
\rho_{41} & \rho_{42} & 0 & 0 
\end{array} \right).  \label{eq:dissipatord}
\end{equation} 
While determining analytically the full eigensystem for the corresponding evolution equation is very involved, 
valuable information can already be obtained by considering a few important eigenvalues. Comparing
this system to the previous case where the dissipative coupling was through the local density, one important
difference is the occurrence of {\it two} eigenvalues with zero values:
\begin{align}
\lambda_{d,0} = 0,  && v_{d,0} = \frac{1}{4}\begin{pmatrix}1 & 0& 0& 0\\ 0& 1& 0& 0\\ 0& 0& 1& 0\\ 0& 0& 0& 1\end{pmatrix}, \\
\lambda'_{d,0} = 0,  && v'_{d,0} = \frac{1}{2}\begin{pmatrix}0& 0& 0& 0\\ 0& 0& 0& 0\\ 0& 0& 1& 1\\ 0& 0& 1& 1\end{pmatrix}. 
\end{align}
This observation implies that the steady state is not unique anymore, but formed from a combination of these two eigenstates. This situation
occurs as the dissipation does not act on singly occupied sites. Therefore, any state in which no doubly occupied sites are present, 
and which is additionally an eigenstate of the kinetic part of the Hamiltonian, is stable. As a consequence a coherence 
between singly occupied states, present in the initial state, survives in the long-time limit. 
In addition to the steady state values, four other eigenvalues also play an important role if the initial state is the ground state of 
the Hamiltonian. These are the roots of the polynomial equation in $\lambda$ 
\begin{eqnarray}
&&24 \gamma^2 \left(\frac{J}{\hbar}\right)^2 + \left[ \gamma^3 +80 \gamma \left(\frac{J}{\hbar}\right)^2+4\gamma \left(\frac{U}{\hbar}\right)^2 \right]\lambda \nonumber\\
&&+ \left[5\gamma^2 +64 \left(\frac{J}{\hbar}\right)^2+4\left(\frac{U}{\hbar}\right)^2 \right]\lambda^2+8\gamma \lambda^3 +4 \lambda^4=0\;. \nonumber \\ 
\end{eqnarray}
In the limit of $J\ll \hbar\gamma,~U$ the corresponding four eigenvalues become
\begin{eqnarray}
&&\lambda_{d,1}=-\gamma, \\
&&\lambda_{d,2}=-\frac{\gamma}{2} -{\rm i}\frac{U}{\hbar},\\
&&\lambda_{d,3}=-\frac{\gamma}{2} +{\rm i}\frac{U}{\hbar},\\
&&\lambda_{d,{\rm Zeno}}=-\frac{24\gamma J^2}{(\hbar\gamma)^2+4U^2}.\\   
\end{eqnarray}
The slowest decay rate is given by the eigenvalue $\lambda_{d,{\rm Zeno}}$. Similarly to $\lambda_{n,{\rm Zeno}}$ for the case of the 
coupling to the local density, $\lambda_{d,{\rm Zeno}}$ presents both the ``Zeno''-behavior proportional to $\frac{J^2}{\hbar^2\gamma}$ 
for large $\gamma$, and an interaction impeding, proportional to $\frac{\gamma J^2}{U^2}$ for strong interaction $U$. It is interesting
to note that the transition to the interaction impeding regime is occurring at smaller values of $U$ compared to $\hbar \gamma$ 
than for $\lambda_{n,{\rm Zeno}}$. Moreover, when the initial state is the ground state of the Hamiltonian, the amplitude of the overlap with this
slowly decaying state increases with increasing $U$. In this regime of large interaction, the slow decaying ``eigenmatrix'' 
is proportional, up to zeroth order in the hopping, to
\begin{equation} 
 v_{d,{\rm Zeno}} = \frac{1}{2}~\begin{pmatrix}-1& 0& 0& 0\\ 0& -1& 0& 0\\ 0& 0& 1& -1\\ 0& 0& -1& 1\end{pmatrix}. 
\end{equation}
This state is clearly invariant under the action of the dissipator $\mathcal{D}_d$, Eq.~(\ref{eq:dissipatord}), 
but is connected via the kinetic term to states which are not invariant under the action of the dissipator. 
Thus, even though the dissipator does not directly act on state $v_{d,{\rm Zeno}}$, this state decays due to the 
interplay of the Hamiltonian and dissipative terms. In contrast, due to a direct coupling to the dissipation, the other 
relevant eigenvalues correspond to much faster decay rates of the order of $\gamma$. Additionally, the imaginary parts 
of $\lambda_{d,2}$ and $\lambda_{d,3}$ induce oscillations (present for large interaction strength), the frequency of which is approximately linear in $U/\hbar$. 

%%%%%%%%%%%%%%%%%%%%%%%%%%%%%%%%%%%%%%%%%%%%%%%%%%%%%%%%%%%%%%%%%%%%%%%%%%%%%
% slave-spin
%

\subsection{Dissipative dynamics within a slave-spin representation approach}\label{sec:slave}

\subsubsection{Definition of the slave-spin representation}

We explain here how the slave-spin representation of fermionic operators, first developed
to study the equilibrium physics of multi-orbital Hubbard models~\cite{deMediciBiermann2005}, can be
successfully applied to explore the short-time dynamics of dissipative fermionic systems. 
As for any slave-field representations, the slave-spin approach relies on
enlarging the Hilbert space and then imposing a local constraint to eliminate unphysical states. 
The objective of this approach is to choose the auxiliary states in a way which will allow one to
handle the model with greater ease. Applying the constraint on average effectively corresponds 
to treating the system within the mean-field approximation. 

The fermionic slave-spin representation was developed following the realization that the two possible occupancies of a spinless fermion 
on a given site, $n_\text{c} = 0$ and $n_\text{c} = 1$, can be considered as the two possible states of a spin-$\frac{1}{2}$ variable, 
$S^z=-\frac{1}{2}$ and $S^z=\frac{1}{2}$, a mapping widely used to represent hard-core bosons. Here the label ``$c$'' denotes the
physical fermion. In the fermionic context, where anticommutation properties need to be maintained, 
one also introduces an auxiliary fermionic field, $f$, and the local constraint $S^z + \frac{1}{2} = f^\dagger f$ which, together, 
translate into a faithful representation of the Hilbert space: 
\begin{eqnarray}
|n_\text{c} &=& 0 \rangle = |n_\text{f} = 0, S^z = -\frac{1}{2} \rangle, \nonumber \\ 
|n_\text{c} &=& 1 \rangle = |n_\text{f} = 1, S^z = +\frac{1}{2} \rangle.
\end{eqnarray}
This representation is often useful as it allows one to choose the auxiliary fermionic sector 
in such a way that it consists solely of non-interacting terms. The more intricate interaction terms 
are all described using auxiliary spins (in other words, in the charge sector) allowing for a more effective
treatment. 

One can correctly wonder why we chose, among the plethora of slave-field representations developed over the years 
for Hamiltonian systems, to extend the slave-spin representation method to dissipative systems.

It is known that different slave-variable representations lead to different 
mean-field theories, and that, generally, the quality of a given mean-field treatment 
can be improved by adapting the slave fields to a system specificities. For equilibrium
problems, one usually tries to find the right balance between the simplicity of the
representation, the number of unphysical states introduced, and the possibility of an 
analytical treatment for the resulting mean-field theory. In the case of time-dependent 
problems, the optimal conditions are, to the best of our knowledge, not as well known. From our experience, 
it appears that a good description of the system dynamics can be achieved if 
the Hilbert space of the relevant sector (charge or spin) is of the correct physical size even when
the constraint is only applied on average. When the size of the sector is enlarged, we noticed that the presence
of spurious states usually lead to unphysical interference effects. For example, we found that the celebrated
slave-rotor approach~\cite{FlorensGeorges2002} cannot be used to study the dynamics of the charge 
sector because, at the mean-field level, the presence of unphysical states leads to catastrophic 
interference effects and thus gives unphysical evolutions.

Within the slave-spin representation, there is a certain freedom in choosing the appropriate
representation for the physical fermions. As we consider here a half-filled
system, we choose the representation
\begin{eqnarray}
\label{eq:ctoSxf}
c_{{\bf r},\sigma}^\dagger &\rightarrow& 2 S^x_{{\bf r},\sigma} f^\dagger_{{\bf r},\sigma}, \nonumber \\
c_{{\bf r},\sigma} &\rightarrow& 2 S^x_{{\bf r},\sigma} f_{{\bf r},\sigma},
\end{eqnarray}
which is correct on the physical Hilbert space. This choice ensures that the single particle spectral weight $Z$ (the amplitude corresponding
to a single particle state) remains equal to one for $U=0$ even when further mean-field approximations are made~\cite{FlorensGeorges2002}. 
With this choice, the Fermi-Hubbard Hamiltonian takes the form
\begin{eqnarray}
H &=& -4 J \!\!\! \sum_{{\bf r}, {\bf r}'({\bf r}),\sigma} \!\!\! f_{{\bf r}\sigma}^\dagger f_{{\bf r}'\sigma} S_{{\bf r}\sigma}^x S_{{\bf r}'\sigma}^x 
+ \frac{U}{2} \sum_{{\bf r}} \left( \sum_\sigma S_{{\bf r}\sigma}^z \right)^2 \nonumber \\
&&+ \;\theta~\sum_{{\bf r},\sigma} \left(S_{{\bf r}\sigma}^z + \frac{1}{2} - f_{{\bf r}\sigma}^\dagger f_{{\bf r}\sigma} \right). 
\end{eqnarray}
Here the last term enforces the local constraint on average using the Lagrange multiplier $\theta$.
As we restrict ourselves to the situation where the system is half-filled, we can set $\theta = 0$. The Lagrange 
multiplier term then disappears all together.

One can further recast the dissipator within the slave-spin representation. However, this rewriting
is not unique as the jump operators can be written using either the auxiliary fermionic or spin 
operators. The most appropriate choice often depends on the tractability of the resulting equation of motion.
We choose here to recast the dissipator, $\mathcal{D}(\rho)$, using exclusively the auxiliary 
spin operators. When the dissipation couples to the local density, the
jump operator is rewritten as $j_{{\bf r}\sigma} = n_{{\bf r}\sigma} \rightarrow S^z_{{\bf r}\sigma} + \frac{1}{2}$
and the dissipator adopts the simple form
\begin{eqnarray}
\mathcal{D}^s_n(\rho) &=& \gamma \sum_{{\bf r},\sigma} \left( S_{{\bf r}\sigma}^z \rho S_{{\bf r}\sigma}^z - \frac{1}{4} \rho \right).
\end{eqnarray}
While for coupling to the double occupancy density, the jump operator is recast as
$j_{{\bf r}\sigma} = \frac{1}{\sqrt{2}} n_{{\bf r}\up}n_{{\bf r}\down} 
\rightarrow \frac{1}{\sqrt{2}} (S^z_{{\bf r}\up} S^z_{{\bf r}\down} 
+ \frac{1}{2} S^z_{{\bf r}\up} + \frac{1}{2} S^z_{{\bf r}\down} + \frac{1}{4})$. In this case, we denote the
dissipator as ${D}^s_d(\rho)$.

\subsubsection{The Lindblad equation within the slave-spin representation}

Now that we have recast the Fermi-Hubbard Hamiltonian and the dissipator within the
slave-spin representation, we can find the corresponding Lindblad equation. We assume 
here that the density matrix representing the system is a direct product of the spin and fermionic 
auxiliary spaces, i.e.~$\rho = \rho^s \otimes \rho^f$. This assumption is often justified but 
implies that the charge and spin sectors are only minimally coupled.
This procedure can be further generalized, if one considers the general decomposition of a density matrix into 
the weighted sum of several such products and evolve each term separately as permitted by the linear character
of the evolution equations. 

Using this choice for the density matrix, the Lindblad equation takes the form
\begin{eqnarray}
&& \dot{\rho}^s \otimes \rho^f + \rho^s \otimes \dot{\rho}^f = \\
&& -\frac{i}{\hbar} \left[ -4 J \!\!\! \sum_{{\bf r}, {\bf r}'({\bf r}),\sigma} \!\!\! f_{{\bf r}\sigma}^\dagger f_{{\bf r}' \sigma} 
S_{{\bf r} \sigma}^x S_{{\bf r}'\sigma}^x \right. \nonumber \\
&& \,\,\,\,\,\,\,\,\,\,\,\,\,\, 
\left. +~ \frac{U}{2} \sum_{\bf r} \left( \sum_\sigma S_{{\bf r}\sigma}^z \right)^2 
, \rho^s \otimes \rho^f \right] +~D^s(\rho^s) \otimes \rho^f \nonumber.
\end{eqnarray}
Taking partial traces over the fermionic and spin auxiliary spaces, we obtain two coupled differential 
equations. The evolution of the auxiliary spin space (the charge sector) is described by the equation
\begin{eqnarray}
\label{eq:dcharge}
\dot{\rho}^s &=&
-\frac{{\rm i}}{\hbar} 
\left[-4J \!\!\! \sum_{{\bf r}, {\bf r}'({\bf r}),\sigma} \!\!\! \text{Tr}_\text{f}\left(f_{{\bf r}\sigma}^\dagger f_{{\bf r}' \sigma}\right)
S_{{\bf r}\sigma}^x S_{{\bf r}'\sigma}^x \right.\\
&& \,\,\,\,\,\,\,\,\,\,\,\,\,\,
\left. +~\frac{U}{2} \sum_{\bf r} \left( \sum_\sigma S_{{\bf r}\sigma}^z \right)^2 , \rho^s \right] + D^s(\rho^s).\nonumber
\end{eqnarray} 
While the evolution of the fermionic part of the density matrix (the spin sector) is given by the differential equation
\begin{eqnarray}
\label{eq:dspin}
\dot{\rho}^f  &=&
-\frac{{\rm i}}{\hbar} \left[-4J \!\!\! \sum_{{\bf r}, {\bf r}'({\bf r}),\sigma} \!\!\! 
\text{Tr}_\text{s} \left(S_{{\bf r}\sigma}^x S_{{\bf r}' \sigma}^x\right) 
f_{{\bf r}\sigma}^\dagger f_{{\bf r}' \sigma}, \rho^f \right]\!\!. 
\end{eqnarray}
To summarize our result up to this point, we now have two coupled mean-field Lindblad equations describing the evolution of the 
spin and charge sectors of a dissipative half-filled fermionic system. Our main approximation was to apply only on average 
the constraint ensuring the correct dimensionality of the system Hilbert space.

\subsubsection{Further mean-field decoupling}

To make further progress, we perform a mean-field decoupling of the charge sector which corresponds to rewriting the density matrix 
as a direct product over all sites: $\rho^s = \bigotimes_{\bf r} \rho^s_{\bf r}$. Additionally, as we are
dealing with a translationally invariant system, we assume that all $\rho^s_{\bf r}$ are equal and denote the
local density matrix as $\tilde{\rho}^s$. Under this simplification, the Lindblad equation for the
auxiliary fermions can be brought to a diagonal form in momentum space:
\begin{eqnarray}
\dot{\rho}^f &=&  -\frac{{\rm i}}{\hbar} \left[ \sum_{{\bf k}\sigma} Z_\sigma(t) ~ \epsilon_{\bf k} ~ n_{{\bf k}\sigma}, \rho^f \right]
\end{eqnarray}
where $Z_\sigma(t)$, the charge spectral weight, is given by 
\begin{eqnarray}
Z_\sigma(t) &=& 4 \left( \text{Tr}_s\!\left(S^x_{\sigma} \tilde{\rho}^s \right) \right)^2 = 4~\langle S^x_\sigma \rangle^2
\end{eqnarray}
with $\epsilon_{\bf k} = -J \sum_{{\bf r}'({\bf r})} e^{{\rm i}~{\bf k}\cdot ({\bf r}-{\bf r}')}$. Interestingly, one can show that when 
the evolution begins from the ground state of the mean-field Fermi-Hubbard model, which corresponds, for the 
auxiliary fermions, to the Fermi sea, the density matrix $\rho^f$ is time independent.

In contrast, the evolution of the charge sector is much more involved as the 
interaction and dissipative terms have been exclusively rewritten using auxiliary spins. 
At the mean-field level, and under the assumption of translational invariance, the 
evolution equation for this sector reads
\begin{eqnarray}
\label{eq:dchargefinal}
\dot{\tilde{\rho}}^s &=&
-\frac{{\rm i}}{\hbar} \left[-J_{\text{eff}}~\sum_{\sigma} \langle S^x_{\sigma} \rangle S^x_{\sigma} + 
\frac{U}{2} \left(\sum_\sigma S^z_{\sigma}\right)^2
,~\tilde{\rho}^s\right] \nonumber \\
&&+~D^s_{\text{loc}}(\tilde{\rho}^s)
\end{eqnarray}
where the effective time-independent auxiliary spin coupling (independent of $\sigma$ at half-filling) is
\begin{eqnarray}
J_\text{eff} = J^\sigma_{\text{eff}} 
&=& - \frac{8}{\Omega} \sum_{\bf k} \epsilon_{\bf k} \text{Tr}_\text{f}\!\left(n_{{\bf k}\sigma} \rho^f\right) \nonumber \\
&=& - \frac{8}{\Omega} \sum_{\bf k} \epsilon_{\bf k} \langle n_{{\bf k}\sigma} \rangle \nonumber
\end{eqnarray}
with $\Omega$ the number of sites. $D^s_{\text{loc}}(\tilde{\rho}^s)$ is the local dissipator defined as
\begin{eqnarray}
D^s_{\text{loc}}(\tilde{\rho}^s) &=&  
\gamma~\sum_{\sigma} \left( j_{\sigma} \tilde{\rho}^s~j_{\sigma} 
- \frac{1}{2} j^2_{\sigma} \tilde{\rho}^s - \frac{1}{2} \tilde{\rho}^s~j^2_{\sigma} \right). \nonumber
\end{eqnarray}
Consequently, to understand the evolution of the charge under both dissipative and
interaction effects, we solve Eq.~(\ref{eq:dchargefinal}) to find $\tilde{\rho}^s(t)$. Alternatively, we can consider directly 
the evolution of any observable of the charge sector, $O$, using the Heisenberg equation~\cite{BreuerPetruccione2002}
\begin{eqnarray}
\label{eq:Heisenberg}
\frac{d}{dt} \langle O \rangle\!\!\! &=& \!\!\!
\left\langle {\rm i} \left[-J_\text{eff}~\sum_{\sigma} \langle S^x_{\sigma} \rangle S^x_{\sigma} 
+ \frac{U}{2} \left(\sum_\sigma S^z_{\sigma}\right)^2, O \right] \right\rangle \nonumber \\
&&+~\langle D^s_{\text{loc}}(O) \rangle
\end{eqnarray}
as $\langle O (t) \rangle = \text{Tr}_\text{s}(O~\tilde{\rho}^s(t))$. As Eq.~(\ref{eq:Heisenberg}) provides
us with a more intuitive understanding of the physics at play, we will predominantly use this alternative
form in the following sections to study the evolution of the charge sector.

\subsubsection{Mean-field ground state in the charge sector}
As mentioned previously, we want to evolve the system from its ground state that it is decoupled
from the dissipative environment. We therefore want to determine the ground state of the
mean-field Hamiltonian
\begin{eqnarray}
H &=& -J_\text{eff} \sum_{\sigma} \langle S^x_{\sigma} \rangle S^x_{\sigma} + 
\frac{U}{2} \left(\sum_\sigma S^z_{\sigma}\right)^2.
\label{eq:MFH}
\end{eqnarray}
As at half-filling $\langle S^u_\sigma \rangle = \langle S^u_{\bar{\sigma}} \rangle$ and
$\langle S^u_{\sigma} S^v_{\bar{\sigma}} \rangle = \langle S^u_{\bar{\sigma}} S^v_{\sigma} \rangle$
for $u,~v \in \{x,y,z\}$, 
we identify the ground state by requiring that 
\begin{eqnarray}
\langle \Psi(J_\text{eff}, U, \langle S^x_{\uparrow} \rangle)~|~S^x_{\uparrow}~|~\Psi(J_\text{eff}, U, \langle S^x_{\uparrow} \rangle) \rangle 
=  \langle S^x_{\uparrow} \rangle
\end{eqnarray}
where $| \Psi(J_\text{eff}, U, \langle S^x_{\uparrow} \rangle) \rangle$ is the ground state wavefunction obtained 
by diagonalizing the mean-field Hamiltonian given by Eq.~(\ref{eq:MFH}) while holding $\langle S^x_{\uparrow} \rangle$ fixed.
Using this self-consistent approach, we find that in the ground state
\begin{eqnarray}
\langle S^x_{\uparrow} \rangle_\text{gs} &=& \frac{1}{4}~\sqrt{4-\left(\frac{U}{J_\text{eff}}\right)^2}
\end{eqnarray}
in the range $0\leq U  \leq 2 J_\text{eff}$, and $0$ for $ U  > 2 J_\text{eff}$. In the first range, 
we also identify that 
\begin{eqnarray}
&& \langle S^x_{\uparrow} S^x_{\downarrow} \rangle_\text{gs} = \frac{1}{4}, \nonumber \\
&& \langle S^y_{\uparrow} S^y_{\downarrow} \rangle_\text{gs} = \frac{U}{8 J_\text{eff}}, \nonumber \\
&& \langle S^z_{\uparrow} S^z_{\downarrow} \rangle_\text{gs} = - \frac{U}{8 J_\text{eff}}, \nonumber
\end{eqnarray}
while 
$\langle S^y_{\uparrow} \rangle$,  
$\langle S^z_{\uparrow} \rangle$,
$\langle S^x_{\uparrow} S^y_{\downarrow} \rangle$, 
$\langle S^x_{\uparrow} S^z_{\downarrow} \rangle$, 
and $\langle S^y_{\uparrow} S^z_{\downarrow} \rangle$ are zero in the ground state. 
Consequently, within this mean-field approximation, we obtain that a metallic phase with
spectral weight $Z_{\sigma} = 1 - \left(\frac{U}{2 J_\text{eff}}\right)^2$ exists for $0\leq U  \leq 2 J_\text{eff}$. 
As expected, our choice of representation for the physical fermions, Eq.~(\ref{eq:ctoSxf}), captures correctly 
that $Z = 1$ at $U = 0$. This simple mean-field model also predicts correctly that the double 
occupation, $\langle n_{\uparrow} n_{\downarrow} \rangle$, is $\frac{1}{4}$ at $U=0$ and zero when the system becomes insulating % at $U = 2 J_\text{eff}$, 
and that the local density, $\langle n_{\uparrow} \rangle + \langle n_{\downarrow} \rangle$, is one for all $U$.
These findings are in agreement with results obtained in previous studies based on the slave-spin method~\cite{deMediciBiermann2005,HassanDeMedici2010}.

\subsubsection{Case (i): Dissipative coupling to the local density}

We can now use our newly developed mean-field formalism to study the evolution of an interacting
fermionic system dissipatively coupled to its environment through the local density,
i.e.~for the jump operator $j_{{\bf r},\sigma} = n_{{\bf r},\sigma}$. We consider here a situation where the fermionic
system is initially metallic (low interaction $U$) and isolated from its environment. As we want to understand how the 
charge spectral weight evolves under the influence of dissipative effects, we need to  
consider the evolution of $\langle S^x_{\sigma} \rangle$. 
We identify a set of four coupled differential equations that needs to be solved together to
uncover the evolution of the spectral weight $Z_{\sigma}$:
\begin{eqnarray}
\label{eq:nsx}
\frac{d}{dt} \langle S^x_{\uparrow} \rangle &=& -\frac{\gamma}{2}~\langle S^x_{\uparrow} \rangle - \frac U {\hbar}~\langle S^y_{\uparrow} S^z_{\downarrow} \rangle, \\
\frac{d}{dt} \langle S^z_{\uparrow} S^z_{\downarrow} \rangle &=& 
- 2~\frac{J_\text{eff}}{\hbar}~\langle S^x_{\uparrow} \rangle \langle S^y_{\uparrow} S^z_{\downarrow} \rangle, \\
\frac{d}{dt} \langle S^y_{\uparrow} S^y_{\downarrow} \rangle &=& 
- \gamma~\langle S^y_{\uparrow} S^y_{\downarrow} \rangle 
+ 2~\frac{J_\text{eff}}{\hbar}~\langle S^x_{\uparrow} \rangle \langle S^y_{\uparrow} S^z_{\downarrow} \rangle, \,\,\, \\
\label{eq:nsysz}
\frac{d}{dt} \langle S^y_{\uparrow} S^z_{\downarrow} \rangle &=& -\frac{\gamma}{2}~\langle S^y_{\uparrow} S^z_{\downarrow} \rangle 
+ \frac{U}{4\hbar}~\langle S^x_{\uparrow} \rangle \nonumber \\
&& +~\frac{J_\text{eff}}{\hbar}~\langle S^x_{\uparrow} \rangle \left( \langle S^z_{\uparrow} S^z_{\downarrow} \rangle - \langle S^y_{\uparrow} S^y_{\downarrow} \rangle \right).
\end{eqnarray}

In the following, we discuss the solution of this set of equations for various instantaneous 
quenches where the metallic state is quenched (at $t=0$) to
analytically solvable limits:

\paragraph{Instantaneous quench to $U=0$.}
When the metallic system is quenched to the non-interacting limit, the equation for $\langle S^x_{\uparrow} \rangle$ is closed and is written as
\begin{eqnarray}
\frac{d}{dt} \langle S^x_{\uparrow} \rangle &=& -\frac{\gamma}{2} \langle S^x_{\uparrow} \rangle,
\end{eqnarray}
which has for solution
\begin{eqnarray}
\langle S^x_{\uparrow}(t) \rangle &=& \langle S^x_{\uparrow} \rangle_{\text{t=0}}~~e^{-\frac{\gamma}{2}t}.
\end{eqnarray}
In this limit, the charge spectral weight, $Z(t) = 4~\langle S^x_{\uparrow}(t) \rangle^2$, decays exponentially 
fast with rate $\gamma$.

\paragraph{Instantaneous quench to $J_{\rm eff} = 0$. }
When the metallic system is quenched to a limit where the lattice sites decouple, considering the two following equations 
is sufficient:
\begin{eqnarray}
\frac{d}{dt} \langle S^x_{\uparrow} \rangle 
&=& -\frac{\gamma}{2} \langle S^x_{\uparrow} \rangle - \frac{U}{\hbar} \langle S^y_{\uparrow} S^z_{\downarrow} \rangle, \nonumber \\
\frac{d}{dt} \langle S^y_{\uparrow} S^z_{\downarrow} \rangle 
&=& -\frac{\gamma}{2} \langle S^y_{\uparrow} S^z_{\downarrow} \rangle 
+ \frac{U}{4{\hbar}} \langle S^x_{\uparrow} \rangle.
\end{eqnarray}
These equations have for solution
\begin{eqnarray}
\langle S^x_{\uparrow}(t) \rangle \!\! &=&  \!\! e^{-\frac{\gamma}{2}t} \left[B_1~\cos \left(\frac{U}{2 \hbar} t\right) 
- 2 B_2~\sin \left( \frac{U}{2 \hbar} t \right) \right]\!\!, \nonumber \\
\langle S^y_{\uparrow} S^z_{\downarrow}(t) \rangle \!\! &=& \!\! 
e^{-\frac{\gamma}{2}t} \left[B_2~\cos \left(\frac{U}{2 \hbar} t\right) + \frac{B_1}{2}~\sin \left(\frac{U}{2 \hbar} t \right)\right]\!\!. \nonumber \\
\end{eqnarray}
As the evolution begins from the metallic state, we obtain that $B_2=0$ and $B_1= \frac{1}{4} \sqrt{4 - \left(\frac{U}{J_\text{eff}}\right)^2}$. 
This result indicates that  $\langle S^x_{\uparrow}(t) \rangle$ decays exponentially with rate $\gamma/2$, 
but that this decay is dressed with oscillations of period $T_p= 4\pi\hbar / U$. 

\subsubsection{Case (ii): Dissipative coupling to the double occupancy}
We also explore the evolution of an interacting fermionic system dissipatively coupled to its environment 
through the local double occupancy, i.e.~for the jump operator $j_{\bf r} = \frac{1}{\sqrt{2}} n_{{\bf r}, \uparrow} n_{{\bf r}, \downarrow}$. 
We consider once again an initial situation where the fermionic system is half-filled, metallic and isolated 
from its environment. In this case, to understand the behavior of the charge spectral weight, one needs to solve the
following set of equations:
\begin{eqnarray}
\frac{d}{dt}\langle S^x_\uparrow \rangle &=& 
- \frac{U}{\hbar} \langle S^y_\uparrow S^z_\downarrow \rangle 
- \frac{\gamma}{2}\left(\langle S^x_\uparrow S^z_\downarrow \rangle + \frac{1}{2} \langle S^x_\uparrow \rangle\right),\,\,\,\,\,\, \label{eq:DOSx} \\
\frac{d}{dt}\langle S^y_\uparrow S^z_\downarrow \rangle 
&=& \frac{U}{4\hbar} \langle S^x_\uparrow \rangle
+ \frac{J_\text{eff}}{\hbar}~\langle S^x_\uparrow \rangle (\langle S^z_\uparrow S^z_\downarrow \rangle - \langle S^y_\uparrow S^y_\downarrow \rangle) \nonumber \\
&&- \frac{\gamma}{4}\left(\langle S^y_\uparrow S^z_\downarrow \rangle + \frac{1}{2} \langle S^y_\uparrow \rangle\right), \\
\frac{d}{dt}\langle S^x_\uparrow S^z_\downarrow \rangle 
&=& - \frac{U}{4\hbar} \langle S^y_\uparrow \rangle
- \frac{J_\text{eff}}{\hbar}~\langle S^x_\uparrow \rangle \langle S^x_\uparrow S^y_\downarrow \rangle \nonumber \\
&&- \frac{\gamma}{4}\left(\langle S^x_\uparrow S^z_\downarrow \rangle + \frac{1}{2} \langle S^x_\uparrow \rangle\right), \label{eq:DOSxSz} \\
\frac{d}{dt}\langle S^y_\uparrow \rangle &=& 
\frac{J_\text{eff}}{\hbar}~\langle S^x_\uparrow \rangle   \langle S^z_\uparrow \rangle 
+ \frac{U}{\hbar} \langle S^x_\uparrow S^z_\downarrow \rangle \nonumber \\
&&- \frac{\gamma}{2}\left(\langle S^y_\uparrow S^z_\downarrow \rangle + \frac{1}{2} \langle S^y_\uparrow \rangle\right), \\
\frac{d}{dt}\langle S^z_\uparrow \rangle &=& 
- \frac{J_\text{eff}}{\hbar}~\langle S^x_\uparrow \rangle   \langle S^y_\uparrow \rangle, \\
\frac{d}{dt}\langle S^x_\uparrow S^x_\downarrow \rangle &=& 
\frac{\gamma}{4} \left(\langle S^y_\uparrow S^y_\downarrow \rangle  -  \langle S^x_\uparrow S^x_\downarrow \rangle\right), \\
\frac{d}{dt}\langle S^x_\uparrow S^y_\downarrow \rangle 
&=& -\frac{\gamma}{2} \langle S^x_\uparrow S^y_\downarrow \rangle
+ \frac{J_\text{eff}}{\hbar}~\langle S^x_\uparrow \rangle \langle S^x_\uparrow S^z_\downarrow \rangle, \label{eq:DOSxSy} \\
\frac{d}{dt}\langle S^y_\uparrow S^y_\downarrow \rangle 
&=& - \frac{\gamma}{4}\left(\langle S^y_\uparrow S^y_\downarrow \rangle - \langle S^x_\uparrow S^x_\downarrow \rangle\right) \nonumber \\
&& + 2~\frac{J_\text{eff}}{\hbar}~\langle S^x_\uparrow \rangle \langle S^y_\uparrow S^z_\downarrow \rangle, \\
\frac{d}{dt}\langle S^z_\uparrow S^z_\downarrow \rangle 
&=& -2~\frac{J_\text{eff}}{\hbar}~\langle S^x_\uparrow \rangle \langle S^y_\uparrow S^z_\downarrow \rangle.
\end{eqnarray}
As for the case of the density coupling, we solve these equations numerically to obtain the behavior of
$\langle S^x_\sigma \rangle$ for general combinations of the parameters $U$, $J_\text{eff}$ and $\gamma$. We present the results for this
general case in a later section. For now we consider the limiting case of an instantaneous quench to the non-interacting
limit $U=0$. In this case, one only needs Eqs.~(\ref{eq:DOSx}), (\ref{eq:DOSxSz}) and (\ref{eq:DOSxSy}) to uncover the
evolution of the spectral weight. After a few algebraic manipulations, these three equations can be combined into one to give
\begin{eqnarray}
\frac{d^2}{dt^2}\langle S^x_\uparrow \rangle  &+& \frac{\gamma}{2} \frac{d}{dt}\langle S^x_\uparrow \rangle  
+ \frac{1}{2} \frac{J_\text{eff}^2}{\hbar^2} \langle S^x_\uparrow \rangle ^3 \nonumber \\
&&- \langle S^x_\uparrow(0) \rangle ^2 \frac{J_\text{eff}^2}{2 \hbar^2} e^{-\frac{\gamma}{2}t} \langle S^x_\uparrow \rangle  = 0
\end{eqnarray}
together with the initial condition
\begin{eqnarray}
\frac{d}{dt}\langle S^x_\uparrow \rangle |_{t=0}  &=& -\frac{\gamma}{2} \left(\langle S^x_\uparrow S^z_\downarrow(0) \rangle + \frac{1}{2} \langle S^x_\uparrow(0) \rangle \right).
\end{eqnarray}
At long times, we find surprisingly that the solution is given by an algebraic decay following
\begin{eqnarray}
\langle S^x_\uparrow(t) \rangle \sim \sqrt{\frac{1}{\gamma~t}}.
\end{eqnarray}
A similar algebraic decay has been seen before for a dissipative coupling to the 
spin density difference~\cite{BernierKollath2013} in fermionic systems, in bosonic 
systems~\cite{PolettiKollath2012,PolettiKollath2013,  WitthautWimberger2011} and in spin systems~\cite{Barthel2013}. However, 
the validity of this result is questionable as this algebraic decay occurs at long times, a regime
where the slave-spin approach combined with the mean-field approximation is believed to lack
accuracy. Obtaining the same result using an alternative method would be highly valuable.

\subsection{Adiabatic elimination method} \label{sec:adiabatic}
%%%%%%%%%%%%%%%%%%%%%%%%%%%%%%%%%%%%%
%%%%%%%%%%%%%%%%%%%%%%%%%%%%%%%%%%%%%
%%%%%%%%%%%%%%%%%%%%%%%%%%%%%%%%%%%%%
%%%%%%%%%%%%%%%%%%%%%%%%%%%%%%%%%%%%%
In this section, we discuss how the adiabatic elimination method is applied to study the evolution of 
strongly-interacting dissipative fermionic systems. Adiabatic elimination is commonly used in 
quantum optics~\cite{GardinerZollerBook, CarmichaelBook,BreuerPetruccione2002} to understand, for example,
the interaction of light fields with single atoms. However, in the area of strongly correlated systems, this method 
has only become popular recently~(see for example \cite{Garcia-RipollCirac2009,SyassenDuerr2008,PolettiKollath2012,
PolettiKollath2013,BernierKollath2013,MuellerZoller2012} and references therein). The idea behind adiabatic elimination is to derive an 
effective description of the {\it slow} degrees of freedom, by coarse graining the time evolution and taking only virtual 
transitions to fast modes into account. Therefore, adiabatic elimination describes well the effective slow (long time) 
dynamics of the system. This approach is complementary to the slave-spin method presented in the previous section as
the latter is good, at the mean-field level, to describe the short time dynamics.

\begin{figure}[t]
\includegraphics[width=\columnwidth]{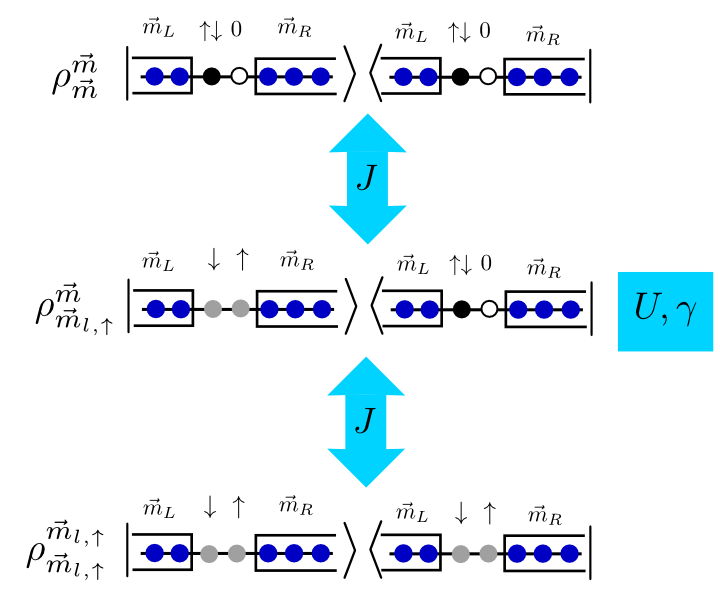}
\caption{(Color online): Schematic of the adiabatic elimination treatment. The slow diagonal 
states $\rho^{\vec{m}}_{\vec{m}}$ and $\rho^{\vec{m}_{l,\uparrow}}_{\vec{m}_{l,\uparrow}}$ are connected 
via excitations (arrow $J$) to the fast decaying state $\rho^{\vec{m}}_{\vec{m}_{l,\uparrow}}$ on 
which both the interaction, $U$, and dissipation, $\gamma$, terms can act. 
\label{fig:adeli}} 
\end{figure}

We concentrate here on case (i): the dissipative coupling with the local density as the quantum jump operator. 
The decoherence free subspace of the dissipator $\mathcal{D}_n$ 
%[see Eq.(\ref{eq:dissipatorn})] 
is given 
by density matrices which are diagonal in the Fock representation. A density matrix in this decoherence free subspace can 
be represented by $\rho=\sum_{\vec{m}}\rho^{\vec{m}}_{\vec{m}}|\vec{m}\rangle\langle\vec{m}|$. Here the 
vector $\vec{m}=(m_{1},m_2,m_{3},\cdots, m_{\Omega})$ denotes a chosen configuration of states. 
The local states on site $j=1,\dots, \Omega$ are denoted by $m_j=0,\su,\sd$ or $\sud$. $\rho^{\vec{m}}_{\vec{m}}$ 
is the weight of the diagonal state: a diagonal element within the Fock representation. Both the unitary evolution, 
caused by the interaction term, and the dissipative evolution leave the diagonal elements of the density matrix $\rho$ invariant. 
In contrast, the action of the dissipator on the off-diagonal terms of a density matrix is proportional to the 
coupling strength $\gamma$ and thus induces a decay at a rate proportional to the coupling strength. Therefore, 
at long times, the dynamics will be dominated by the diagonal terms. In this regime, the evolution is driven by 
the kinetic term of the Hamiltonian which couples the decoherence free subspace to small and fast decaying 
off-diagonal terms (see Fig.~\ref{fig:adeli}). Coarse graining the time-scale of the dynamical evolution allows one 
to describe the system effective slow dynamics using solely diagonal matrices as the connection to the 
off-diagonal matrices is integrated out. This effective dynamics, which plays out within the decoherence free subspace, can be formulated as a 
classical master equation for the different diagonal configurations as derived below.

In the following, we will first show an example of how the integration of the fast degrees of freedom is performed 
in order to obtain the classical master equations. Afterwards, we will perform the mean-field decoupling, and derive 
an effective equation for the dynamics of the reduced single site density matrix. For notational simplicity we will
refer here to a linear chain, however, these results can be easily extended to higher dimensions.

\subsubsection{Adiabatic elimination of the fast modes}   

As an illustrative example, we will derive in the following the classical master equations determining the evolution 
of the weight $\rho^{\vec{m}}_{\vec{m}}(t)$ corresponding to a configuration of atoms in which the site $l$ is 
doubly occupied, i.e.~$m_l=\sud$. 

Off-diagonal elements are generated starting from the diagonal configuration by the application of a 
local hopping term, e.g.~$\rho^{\vec{m}}_{\vec{m}_{l,\sigma}} |\vec{m}_{l,\sigma}\rangle\langle\vec{m}| 
= \left(c^{\dagger}_{l+1,\sigma}c_{l,\sigma}\;\rho^{\vec{m}}_{\vec{m}}|\vec{m}\rangle\langle\vec{m}|\right)$ 
with $\sigma=\su,\;\sd$. The action is only non-zero for $m_{l+1}=0$ or $m_{l+1}=\bar{\sigma}$, where $\bar{\sigma}$ 
is the opposite of $\sigma$. We will only discuss these cases in the following. The corresponding generated 
configuration is then given by $\vec{m}_{l,\sigma}=(\vec{m}_{L},m'_{l,\sigma},m'_{l+1,\sigma},\vec{m}_{R})$ 
with $m'_{l,\sigma}=\bar{\sigma}$ and $m'_{l+1,\sigma}=\sigma$ for  $m_{l+1}=0$ and $m'_{l+1,\sigma}=\sud$ for $m_{l+1}=\bar{\sigma}$. 
$\vec{m}_{L}$ and $\vec{m}_{R}$ are, respectively, the vector of the configuration of fermions to the left of 
site $l$ and to the right of site $l+1$. Neglecting small and fast decaying off-diagonal terms, the evolution of $\rho^{\vec{m}}_{\vec{m}_{l,\sigma}}$ is given by 
\begin{eqnarray} 
\hbar\partial_t \rho^{\vec{m}}_{\vec{m}_{l,\sigma}} &=& \left[-\hbar\gamma + {\rm i} U \left(1-n_{l+1,\bar{\sigma}}\right) \right] \rho^{\vec{m}}_{\vec{m}_{l,\sigma}} \nonumber \\ 
&&+ {\rm i}  J \left(\rho^{\vec{m}}_{\vec{m}} - \rho^{\vec{m}_{l,\sigma}}_{\vec{m}_{l,\sigma}} \right),       
\end{eqnarray}
where $n_{l+1,\bar{\sigma}}$ is the occupation of the particle with spin $\bar{\sigma}$ on site $l+1$ in the configuration $\vec{m}$.
This equation can be integrated giving~\cite{SaitoKayanuma2002}  
\begin{eqnarray} 
\rho^{\vec{m}}_{\vec{m}_{l,\sigma}}(t) &=&  e^{-\left(\gamma-{\rm i} U(1-n_{l+1,\bar{\sigma}})/\hbar\right) t} \rho^{\vec{m}}_{\vec{m}_{l,\sigma}}(0) \nonumber\\ 
&+& {\rm i}  (J/\hbar) e^{-\left(\gamma-{\rm i} U(1-n_{l+1,\bar{\sigma}})/\hbar\right) t} \nonumber \\ 
&\times&\!\!\int_0^{t}\!\!e^{\left(\gamma-{\rm i} U(1-n_{l+1,\bar{\sigma}})/\hbar\right) \tau}
\left(\rho^{\vec{m}}_{\vec{m}}(\tau) - \rho^{\vec{m}_{l,\sigma}}_{\vec{m}_{l,\sigma}}(\tau) \right)d\tau.  \nonumber\\       
\end{eqnarray}      
After an integration by parts and neglecting the fast decaying and dephasing terms, this leads to  
\begin{eqnarray} 
\rho^{\vec{m}}_{\vec{m}_{l,\sigma}}(t)=  -\frac{J}{U(1-n_{l+1,\bar{\sigma}})+{\rm i}\hbar\gamma} 
\left(\rho^{\vec{m}}_{\vec{m}}(t) - \rho^{\vec{m}_{l,\sigma}}_{\vec{m}_{l,\sigma}}(t) \right). \nonumber \\ \label{eq:A0}        
\end{eqnarray}      
The diagonal density matrices evolution is only connected to these off-diagonal elements thus giving  
\begin{eqnarray}
\frac{d}{dt} \rho^{\vec{m}}_{\vec{m}} &=& \frac{J}{\hbar} \sum_{l,\sigma}\left( {\rm i} \rho^{\vec{m}}_{\vec{m}_{l,\sigma}}  + {\rm h.c.} \right).   \label{eq:rhoev0}
\end{eqnarray}     
Inserting Eq.~(\ref{eq:A0}) into Eq.~(\ref{eq:rhoev0}), we obtain
\begin{eqnarray}
\frac{d}{dt} \rho^{\vec{m}}_{\vec{m}} &=& - \sum_{l,\sigma} \frac{J^2\gamma}{(\hbar\gamma)^2+U^2(1-n_{l+1,\bar{\sigma}})^2} 
\left( \rho^{\vec{m}}_{\vec{m}} - \rho^{\vec{m}_{l,\sigma}}_{\vec{m}_{l,\sigma}}  \right).  \nonumber \\ \label{eq:rhoev1}
\end{eqnarray} 
For the general case, where $m_l$ and $m_{l+1}$ can take any value, the evolution equation naturally becomes
\begin{eqnarray}
\frac{d}{dt} \rho^{\vec{m}}_{\vec{m}}\!\!\! &=&\!\!\! - \sum_{l,\sigma} \frac{J^2\gamma\left[n_{l+1,\sigma}(1-n_{l+1,\sigma})\right]}{(\hbar\gamma)^2+U^2(n_{l,\bar{\sigma}}-n_{l+1,\bar{\sigma}})^2} 
\left( \rho^{\vec{m}}_{\vec{m}} - \rho^{\vec{m}_{l,\sigma}}_{\vec{m}_{l,\sigma}}  \right).  \nonumber \\ \label{eq:rhoev01}
\end{eqnarray} 
We have therefore derived the effective classical master equations describing the evolution of the diagonal elements 
of the density matrices by integrating out the off-diagonal parts. 

\subsubsection{Thermodynamic limit}   

The set of equations (\ref{eq:rhoev1}) can in principle be solved numerically, i.e.~by classical Monte-Carlo methods. 
However, as a large number of degrees of freedom is still involved, we further simplify the equations by 
considering the thermodynamic limit. We take the number of atoms of spin $\sigma$, $N_{\sigma}$, to infinity while 
keeping $\nb_{\sigma}$ constant. In this case, a useful approach is to use the factorized 
ansatz $\rho=\bigotimes_l\left[\sum_{m}\rho_{m,l}(t)|m\rangle\langle m|\right]$, 
where $l$ labels the different sites. We further consider a translationally invariant system,  
and for this reason we will drop the $l$ index in the following. The reduced single site density 
matrix $\tilde{\rho}=\sum_{m}\tilde{\rho}_{m}(t)|m\rangle\langle m|$ has only one degree of freedom 
which we chose as the element $\tilde{\rho}_{\sud}(t)$. The other elements follow from the knowledge of the filling 
and from the trace of $\tilde{\rho}$ being unity. The elements of $\tilde{\rho}$ are thus given by       
\begin{subequations} \label{eq:relationdiag}   
\begin{align}
 \tilde{\rho}_{0}(t)&=1-\nb_{\su}-\nb_{\sd}+\tilde{\rho}_{\sud}(t), \\ 
 \tilde{\rho}_{\su}(t)&=\nb_{\sd}-\tilde{\rho}_{\sud}(t), \\ 
 \tilde{\rho}_{\sd}(t)&=\nb_{\su}-\tilde{\rho}_{\sud}(t).   
\end{align}
\end{subequations}
Using Eq.~(\ref{eq:rhoev1}), the time evolution of $\tilde{\rho}_{\sud}$ becomes 
\begin{eqnarray} 
\frac{d\tilde{\rho}_{\sud}}{dt}&=&z\frac{d}{dt}  
\left( \tilde{\rho}_{\sud}\tilde{\rho}_{0}+ \tilde{\rho}_{\sud}\tilde{\rho}_{\su} + 
\tilde{\rho}_{\sud}\tilde{\rho}_{\sd} + \tilde{\rho}_{\sud}\tilde{\rho}_{\sud}\right) \nonumber\\ 
&=& -\left[\frac{2J^2\gamma}{(\hbar\gamma)^2+U^2}\left( 2 \tilde{\rho}_{\sud}\tilde{\rho}_{0} - 
\tilde{\rho}_{\sd}\tilde{\rho}_{\su} - \tilde{\rho}_{\su}\tilde{\rho}_{\sd}  \right)    \right. \nonumber \\ 
&& \left. +\frac{2J^2}{\hbar\gamma^2}\left( \sum_{\sigma=\su,\sd} \tilde{\rho}_{\sud}\tilde{\rho}_{\sigma} - 
\tilde{\rho}_{\sigma}\tilde{\rho}_{\sud}  \right)  \right] \nonumber \\
&=& -\frac{4zJ^2\gamma}{(\hbar\gamma)^2+U^2} \left(  \tilde{\rho}_{\sud}\tilde{\rho}_{0} - 
\tilde{\rho}_{\sd}\tilde{\rho}_{\su}  \right)    
\end{eqnarray}
where $z$ is the lattice connectivity. 
Using Eqs.~(\ref{eq:relationdiag}) we thus get     
\begin{eqnarray}
\frac{d\tilde{\rho}_{\sud}(t)}{dt} = \gad\left(\nb_{\su}\nb_{\sd}-\tilde{\rho}_{\sud}(t)\right)  \label{eq:adfermloc}   
\end{eqnarray}
where $\gad=\frac{8J^2\gamma}{\hbar^2\gamma^2+U^2}$ for a one-dimensional lattice. It is worth noting the 
similarity between $\gad$ and $\gzeno$. Eq.~(\ref{eq:adfermloc}) is, as above expressed, consistent with the 
analytically derived steady state (Eqs.~(\ref{eq:ss1}) through (\ref{eq:ss4})), and
is easily integrated to give 
\begin{equation}
\tilde{\rho}_{\sud}(t)=\nb_{\su}\nb_{\sd}+(\tilde{\rho}_{\sud}(0)-\nb_{\su}\nb_{\sd})e^{-\gad t}. 
\end{equation} 
An exponential decay with a suppressed rate $\gad$ is found which is in agreement with the two site result. 
As discussed for the two site case, the Zeno-effect induces a decreasing decay rate of $J^2/(\hbar^2\gamma)$ 
for large bare coupling strength $\gamma$. Additionally, the interaction is found to slow down considerably the 
decay rate as $\gamma J^2/U^2$ for large interaction strength. A similar dynamical effect had been observed 
in dissipative bosonic systems~\cite{PichlerZoller2010,PolettiKollath2012,PolettiKollath2013,ShchesnovishKonotop2010, BrazhnyiOtt2009, WitthautWimberger2011}.

We can further use the dynamics of the diagonal terms, to compute the coherence of the systems using Eq.~(\ref{eq:A0}). 
The time evolution of the coherence is given by 
\begin{eqnarray}
 C(t)&=&-\frac{8JU}{\hbar^2\gamma^2+U^2}   \left(  \tilde{\rho}_{\sud}\tilde{\rho}_{0} - \tilde{\rho}_{\sd}\tilde{\rho}_{\su} \right) \nonumber \\       
 &=&-\frac{8JU}{\hbar^2\gamma^2+U^2} (\tilde{\rho}_{\sud}(0)-\nb_{\su}\nb_{\sd})e^{-\gad t},     
\end{eqnarray}
whereas the evolution equation for the fluctuations is given by 
\begin{eqnarray}
 \kappa(t)&=&\left( \tilde{\rho}_{\sd}(t)+\tilde{\rho}_{\su}(t)+4 \tilde{\rho}_{\sud}(t)   \right) - (\nb_{\su}+\nb_{\sd})^2 \nonumber \\       
 &=& \left\{ \nb_{\su}+\nb_{\sd} +2\left[\nb_{\su}\nb_{\sd}+(\tilde{\rho}_{\sud}(0)-\nb_{\su}\nb_{\sd})e^{-\gad t}\right]  \right\} \nonumber \\ 
&&- (\nb_{\su}+\nb_{\sd})^2 \nonumber \\ 
 &=&  \kappa(\infty) +2\left(\tilde{\rho}_{\sud}(0)-\nb_{\su}\nb_{\sd}\right)  e^{-\gad t}         
\end{eqnarray}
The presence of strong interaction can thus cause the dynamics to dramatically slow down as the characteristic exponent is inversely 
proportional to $U^2$.

%%%%%%%%%%%%%%%%%%%
\section{Dissipative evolution}\label{sec:dissipativeevolution}
\subsection{Dissipative coupling to the local density}\label{sec:density}

Dissipative coupling to the local density causes the fermionic system to heat up and to flow, at
long times, towards the infinite temperature state, i.e. the totally mixed state proportional to unity in the Fock basis. 
Even though this final state is fairly simple, the evolution towards this state is very interesting
and presents various features which depend on the interplay between the interaction, 
dissipation and hopping terms. We focus here on the evolution of the local kinetic energy as many of the
interesting dynamical features are exemplified by this quantity. Within the slave-spin representation, 
at the mean-field level, the local kinetic energy is given by 
$E_\text{kin}(t) = - J~\sum_\sigma~\aver{c^\dagger_{j,\sigma} c_{j',\sigma}} 
= -J_\text{eff}~\sum_\sigma \aver{S^x_\sigma(t)}^2$, where $j$ and $j'$ are neighboring sites. 
Whereas for the two fermions on two sites toy model, the kinetic energy is
$E_\text{kin}(t) = -2~J~\text{Re}\left[\rho_{13} + \rho_{23} - \rho_{14} - \rho_{24} \right]$ where the $\rho_{ll'}$ 
are entries in the $4 \times 4$ density matrix representing the system. For all situations considered here, 
we choose the initial state to be the ground state of the metallic phase, consequently $E_\text{kin}(0) \neq 0$. 
Such a phase exists in the range $0 < U < 2~J_\text{eff}$ when the $d$-dimensional fermionic system 
is represented within the mean-field slave-spin 
approach. In contrast, for the two fermions on two sites model, the kinetic energy is finite for 
all finite values of $U/J$. 

\begin{figure}[t]
\includegraphics[width=\columnwidth]{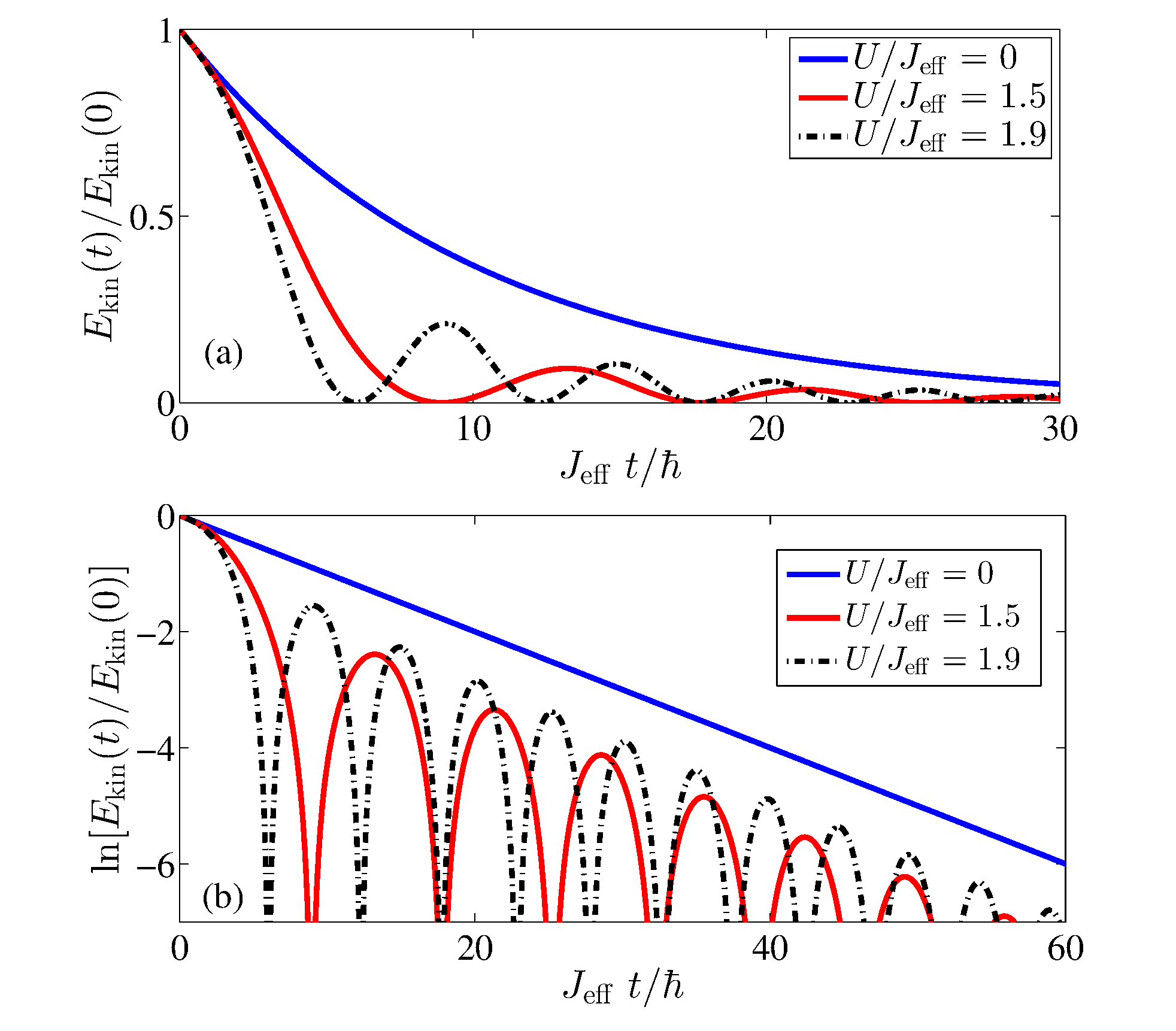}
\caption{(Color online): Time-evolution of the normalized kinetic energy 
for a dissipative coupling $\hbar\gamma = 0.1 J_{\text{eff}}$ to the local density 
and for various interaction strengths in (a) linear scale and (b) log scale. 
These results are obtained within the slave-spin approach solving Eqs.~(\ref{eq:nsx}) to (\ref{eq:nsysz}). The evolution begins
from the metallic ground state of the mean-field Hamiltonian describing the 
charge sector (Eq.~(\ref{eq:MFH})) with $Z(t=0) = 1,~0.44,~0.098$ for $U/J_{\text{eff}} = 0,~1.5,~1.9$ respectively.
\label{fig:sx_density_g}} 
\end{figure}

% non-interacting limit
When the evolution of the metallic state takes place in the absence of interaction, i.e. at $U=0$, the kinetic 
energy decays exponentially as $E_\text{kin}(t) = E_\text{kin}(0)~e^{-\gamma t}$. Within the slave-spin approach, using
the Heisenberg picture for the master equation, this result is recovered and is illustrated in 
Fig.~\ref{fig:sx_density_g}.

\begin{figure}[t]
\includegraphics[width=0.7\columnwidth]{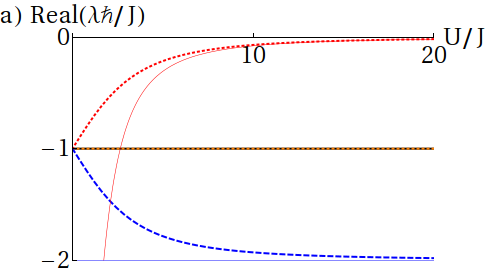}
\includegraphics[width=0.7\columnwidth]{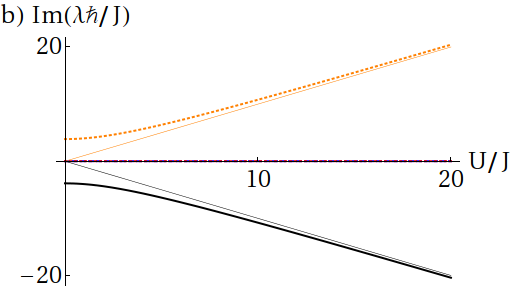}
\caption{(Color online) Real (a) and imaginary (b) parts for $\lambda_{n,2}$ (solid black), 
$\lambda_{n,3}$ (dotted orange), $\lambda_{n,4}$ (dashed blue) and 
$\lambda_{n,{\rm Zeno}}$ (dotted red) versus $U/J$ for a system of two fermions 
(one spin up and one spin down) on two sites with dissipative coupling $\hbar\gamma/J = 1$
to the local density. The solid thin lines correspond to analytical predictions for the 
real parts $-8\gamma J^2/[(\hbar\gamma)^2+U^2]$ (red) and $-2\gamma$ (blue), and 
for the imaginary parts $ U/\hbar$ (orange) and $-U/\hbar$ (black). The real parts 
of $\lambda_{n,2}$ and $\lambda_{n,3}$ are both equal to $-J/\hbar$ whereas
the imaginary parts of both $\lambda_{n,4}$ and $\lambda_{n,{\rm Zeno}}$ are null.
\label{fig:eigenvalues_density_g}} 
\end{figure}

%interaction, oscillations
The presence of interaction alters this behavior. Typical evolutions of the kinetic energy for different interaction 
strengths, evaluated within the slave-spin approach, are illustrated in Fig.~\ref{fig:sx_density_g}. As it can be seen on this 
figure, in addition to the exponential decay, the presence of interaction causes oscillations. For large interaction 
strengths, $U \gg \hbar \gamma,~J_{\rm eff}$, the oscillation period is approximately inversely proportional to 
the interaction, i.e. $T_p \approx \frac{4\pi \hbar}{U}$. This finding is supported by an analysis of the behavior of 
the eigenvalues extracted from the two-fermion, two-site system (see section \ref{sec:2sites2fermionsdensity}). 
Whereas eigenvalues $\lambda_{n,0}$, $\lambda_{n,1}$, $\lambda_{n,4}$ and $\lambda_{n,\text{Zeno}}$ are purely real and 
therefore only cause an exponential decay, the imaginary parts of both $\lambda_{n,2}$ and $\lambda_{n,3}$ 
are proportional to $\pm U/\hbar$ in the large $U$ limit (see Fig.~\ref{fig:eigenvalues_density_g}). 

\begin{figure}[t]
\includegraphics[width=\columnwidth]{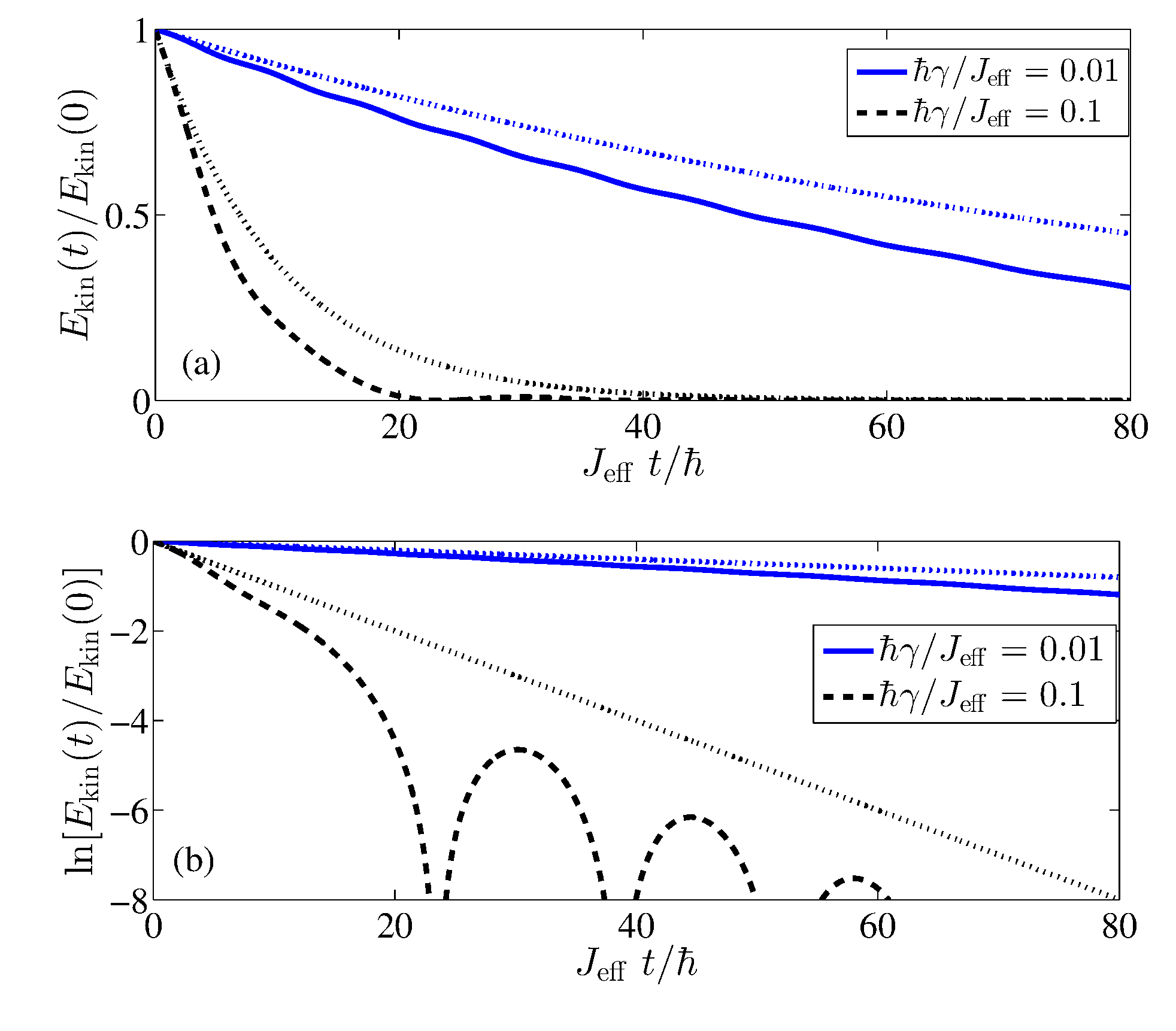}
\caption{(Color online): Time-evolution of the normalized kinetic energy for 
different dissipative couplings to the local density for an interaction 
strength $U = J_{\text{eff}}$ in (a) linear scale and (b) log scale. 
The results are obtained within the slave-spin approach solving Eqs.~(\ref{eq:nsx}) to (\ref{eq:nsysz}). 
Here again the evolution begins from the metallic ground state of the mean-field Hamiltonian 
describing the charge sector (Eq.~(\ref{eq:MFH})) with $Z(t=0) = 0.75$ for $U = J_{\text{eff}}$.
The two dotted lines denote pure exponential decay $\propto e^{-\gamma t}$ for
$\hbar \gamma /J_{\text{eff}} = 0.01$ and $0.1$.
\label{fig:sx_density_U}} 
\end{figure}

\begin{figure}[t]
\includegraphics[width=0.75\columnwidth]{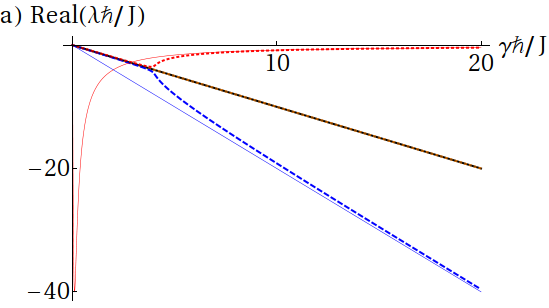}
\includegraphics[width=0.75\columnwidth]{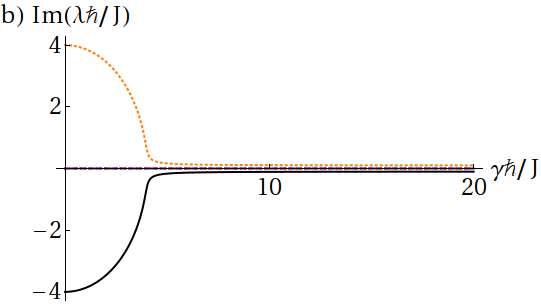}
\caption{(Color online): Real (a) and imaginary (b) parts for $\lambda_{n,2}$ (solid black), 
$\lambda_{n,3}$ (dotted orange), $\lambda_{n,4}$ (dashed blue) and 
$\lambda_{n,{\rm Zeno}}$ (dotted red) versus $\hbar\gamma/J$ for a system of two fermions 
(one spin up and one spin down) on two sites with interaction strength $U/J=0.1$ and 
dissipative coupling to the local density. The solid thin lines correspond to analytical predictions for the real 
parts $-8\gamma J^2/[(\hbar\gamma)^2+U^2]$ (red) and $-2\gamma$ (blue). 
The imaginary parts of both $\lambda_{n,4}$ and $\lambda_{n,{\rm Zeno}}$ are null.
\label{fig:eigenvalues_density_U}} 
\end{figure}

In the opposite limit, for small interaction strengths, $U \ll \hbar\gamma,~J$, the imaginary parts 
of $\lambda_{n,2}$ and $\lambda_{n,3}$ exhibit a different behavior. To linear order in $U$, we find that the imaginary 
parts go as $\sim  \mp 4 J/\hbar$ for $4J \gg \hbar \gamma,~U$ while $\sim \mp U/\hbar$ for $\hbar\gamma \gg 4J,~U$. 
The behavior of both $\lambda_{n,2}$ and $\lambda_{n,3}$ is illustrated in Fig.~\ref{fig:eigenvalues_density_U}: 
for $U = 0.1 J$ and $\hbar \gamma \gg 4J$ the imaginary parts of these eigenvalues are approximately 
equal to $\pm U/\hbar$, whereas for $\hbar \gamma < 4J$ their imaginary parts grow rapidly to $\pm 4J/\hbar$. 
For a system made of many sites, the dependence of the oscillation frequency of the kinetic energy with 
increasing $\gamma$ is seen in Fig.~\ref{fig:sx_density_U} (result obtained
using the slave-spin approach). At low $\gamma=0.01\hbar/J_{\rm eff}$, fast oscillations (with a weak amplitude) can be seen 
in Fig.~\ref{fig:sx_density_U} (a) compared to slower oscillations evident in the logarithmic 
scale in  Fig.~\ref{fig:sx_density_U} (b) for $\gamma=0.1\hbar/J$. 
This figure also shows that in an extended systems, at low interaction strength,
the exponential decay of the kinetic energy, $E_\text{kin}(t) \propto e^{-\gamma t}$, exact for $U=0$, still holds 
for the oscillation envelope. 

However, in Fig.~\ref{fig:sx_density_U}, an initial decay rate slightly 
larger than $\gamma$ can be seen. To understand this deviation, we analyze the structure of the real parts of the eigenvalues 
for the system of two fermions on two sites. We find that the real parts of eigenvalues  $\lambda_{n,2}$ 
and $\lambda_{n,3}$ are approximately $-\gamma$. However, the real part of $\lambda_{n,1}$ and, for large $\gamma$, 
the real part of $\lambda_{n,4}$ both take value close to $-2\gamma$. This is clearly 
depicted in Fig.~\ref{fig:eigenvalues_density_U}. 
 
\begin{figure}[t]
\includegraphics[width=\columnwidth]{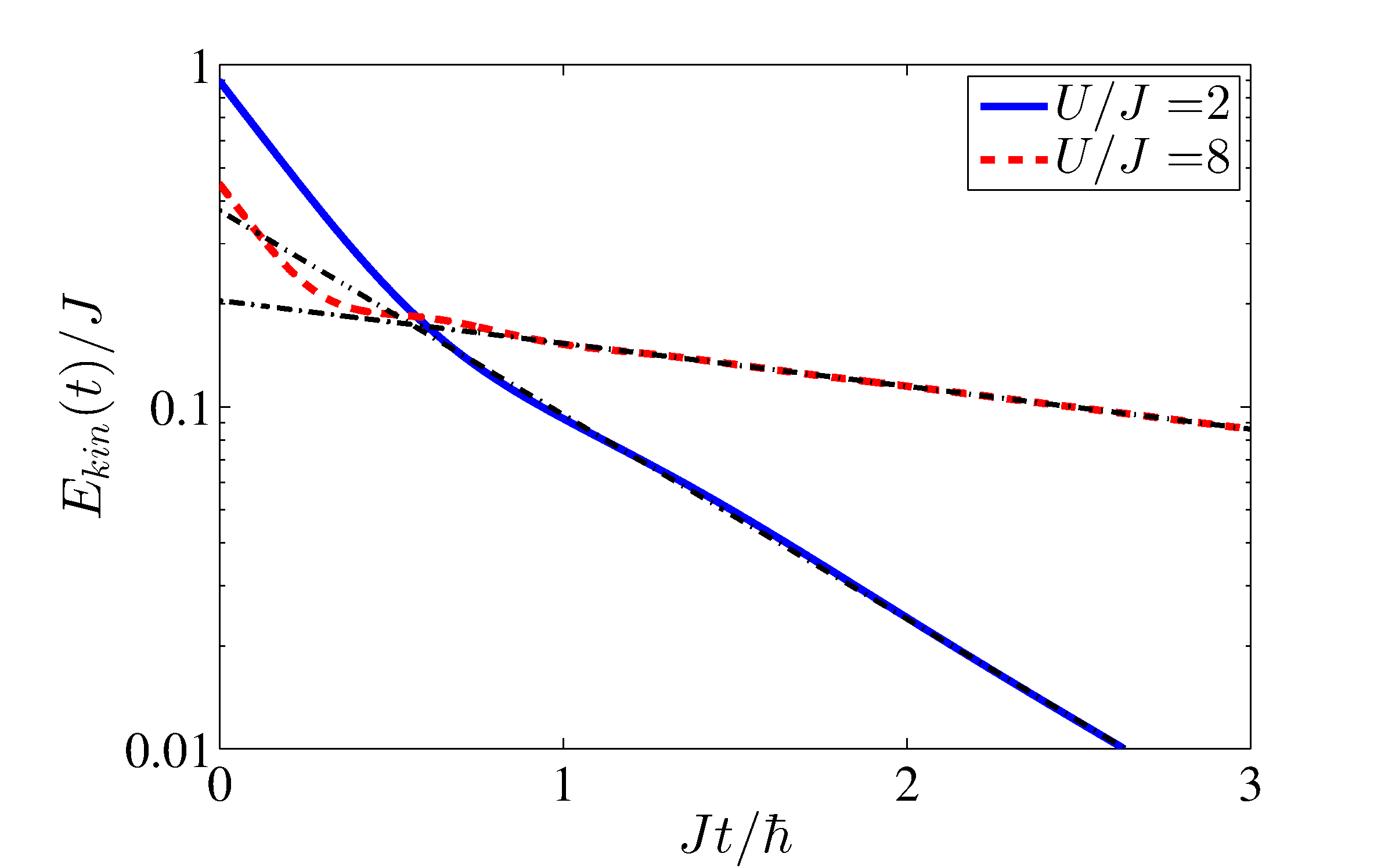}
\caption{(Color online): Time-evolution of the kinetic energy for a two-site two-fermion system 
with a dissipative coupling $\hbar \gamma/J=3$ to the local density. This figure highlights the
presence of two regimes: a short time exponential decay with a rate proportional to $\gamma$ and 
a long time decay dominated by the rate $\lambda_{\text{n, Zeno}}$. The black dash-dotted lines are 
the corresponding Zeno predictions. 
\label{fig:Ekin_density_ed}} 
\end{figure}

%decay rate and Zeno
The behavior of the real part of $\lambda_{n,\text{Zeno}}$ is even more fascinating: for $\hbar\gamma,~U \gg J$, 
this eigenvalue approaches zero as $\gzeno = -\frac{8 \gamma J^2}{(\hbar\gamma)^2+U^2}$ (see Figs.~\ref{fig:eigenvalues_density_g} 
and \ref{fig:eigenvalues_density_U}). This result implies that the components of the density matrix which overlap with 
the corresponding eigenvector become very stable and the system evolves very slowly. Therefore, the system dynamics exhibits 
two regimes: a short time exponential decay with a rate proportional to $\gamma$ and a long time decay 
dominated by the rate $\gzeno$. In Fig.~\ref{fig:Ekin_density_ed}, the presence of these two regimes can be seen 
in the time-evolution of the kinetic energy for the two-site system.

The long time exponential decay of the kinetic energy with rate $\gzeno$ is also recovered in the extended 
system when its evolution is analyzed using the adiabatic elimination technique (see section~\ref{sec:adiabatic}). 
Within this approach, a simple picture for the origin of this effect is gained by interpreting the dissipative coupling 
as a measurement process. A large dissipative coupling acts as a projection onto the dissipation free subspace where 
no evolution occurs. Therefore, at long times, the state evolution is slowed down by the dissipative measurement 
as it is inevitably close to the dissipation free subspace. In contrast to the usual Zeno picture which also occurs in absence of interaction, in the 
fermionic systems under study 
an additional process occurs due to the presence of strong interactions. For large interaction strengths, the decay 
rate is suppressed as $\gamma J^2/U^2$. This strong suppression of the decay rate can be understood by 
considering the balance of energy. Decoherence is caused by the proliferation of excitations, a process requiring a 
large amount of (interaction) energy. As this energy can only be injected into the system by the dissipative process itself,
very few excitations are generated.

%failure of slaves for Zeno
Within the slave-spin approach, only a weak dependence of the decay rate on the interaction strength is 
observed (Fig.~\ref{fig:sx_density_g}) and the Zeno effect is not visible. We attribute this situation to the single-site 
mean-field decoupling used to study the system evolution within the slave-spin approach. 
Due to this mean-field decoupling, $\langle S^x_\uparrow \rangle$ decays rapidly for large 
dissipative couplings to the local density: the connection to the bath of sites is therefore suppressed and the 
system freezes artificially. The use of such mean-field approximations can thus result in the appearance 
of artificial steady states that are not the true long time limits, consequently these approximations must be 
used with great care. However, the absence of the Zeno effect in this particular case is not a fundamental flaw of the 
slave-spin approach as it is due to our drastic mean-field decoupling. In fact, we will see in the next section that, 
for a dissipative coupling to the double occupation, the Zeno effect is recovered within the slave-spin 
approach even at the level of the single-site mean-field decoupling.
 
\subsection{Dissipative coupling to the local double occupancy}\label{sec:doublons}

Similarly to the coupling to the local density, a dissipative coupling to the local 
double occupancy also causes the fermionic system to heat up. However, in the latter case, the steady 
state occurring in the long time limit depends crucially on the initial state. In particular, we find that 
the single particle coherence can survive even under the presence of dissipation, and that the final state is 
a combination of the infinite temperature state and of eigenstates of the kinetic part of the Hamiltonian. This 
eigenstate possesses this special feature that it is only made of singly occupied sites.

\begin{figure}[t]
\includegraphics[width=\columnwidth]{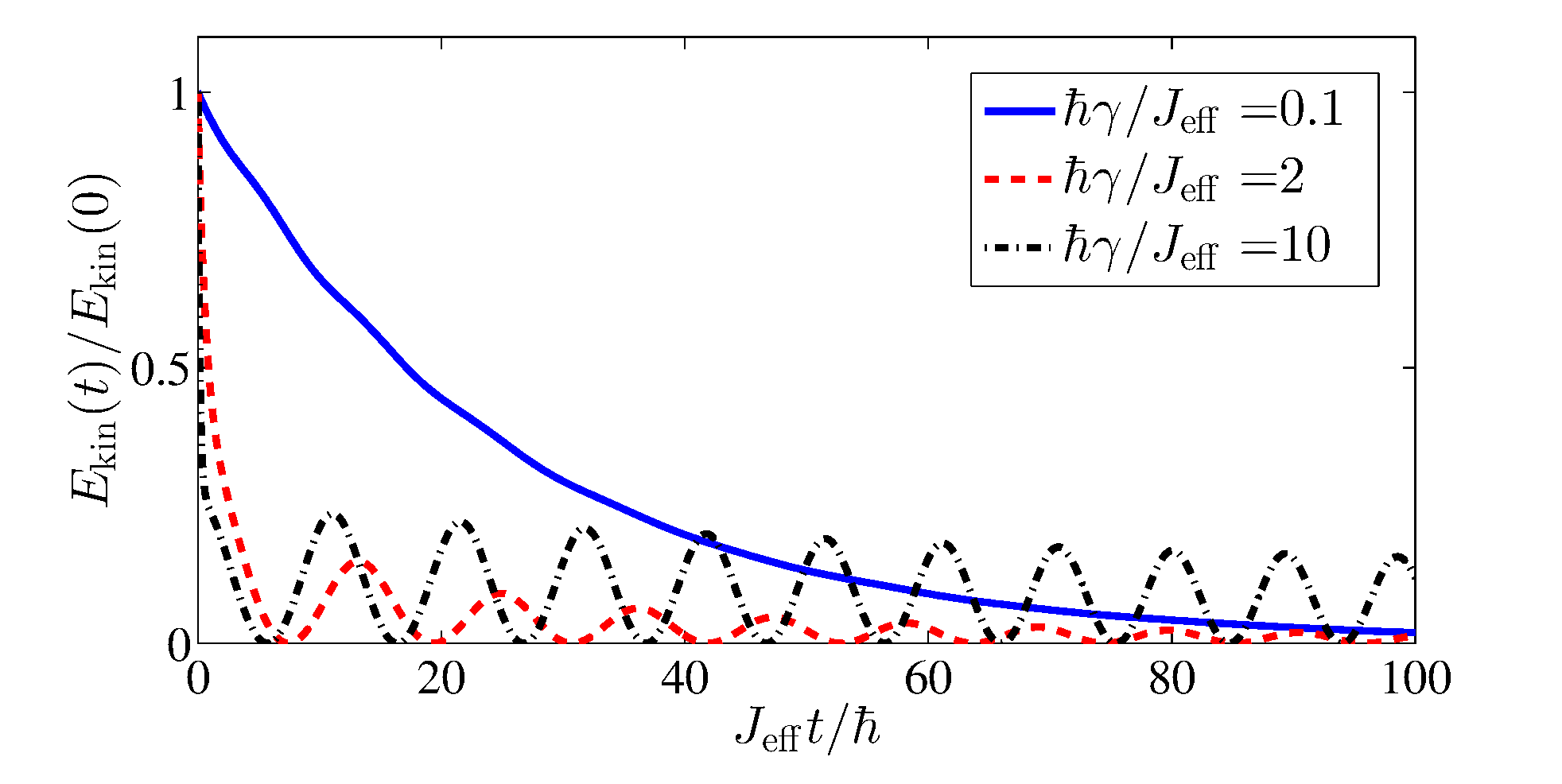}
\caption{(Color online): Time-evolution of the normalized kinetic energy for an 
interaction strength $U = J_{\text{eff}}$ and different dissipative couplings to the local double occupancy.
The results are obtained within the slave-spin approach solving Eqs.~(\ref{eq:DOSx}) to (\ref{eq:DOSxSy}).
The evolution begins from the metallic ground state of the mean-field Hamiltonian describing the 
charge sector (Eq.~(\ref{eq:MFH})) with $Z(t=0) = 0.75$ for $U = J_{\text{eff}}$.
\label{fig:sx_doublon_U}} 
\end{figure}

\begin{figure}[t]
\includegraphics[width=0.7\columnwidth]{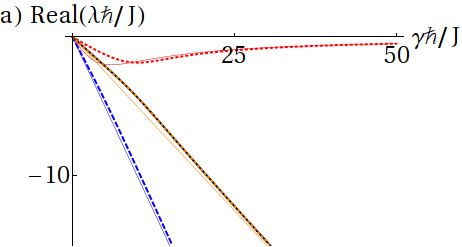}
\includegraphics[width=0.7\columnwidth]{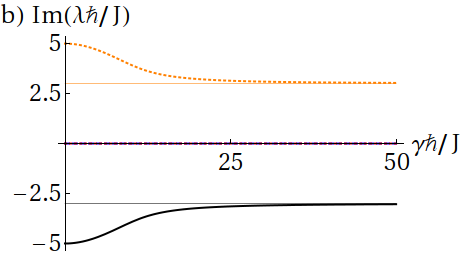}
\caption{(Color online): Real (a) and imaginary (b) parts for $\lambda_{d,1}$ (dashed blue), 
$\lambda_{d,2}$ (black), $\lambda_{d,3}$ (dotted orange) and $\lambda_{n,{\rm Zeno}}$ (dotted red) 
versus $\hbar\gamma/J$ for two fermions (one spin up and one spin down) on two sites with $U/J=3$ and dissipative
coupling to the local double occupancy. The solid thin lines correspond to analytical predictions for the real 
parts $-24\gamma J^2/[(\hbar\gamma)^2+U^2]$ (red), $-\gamma/2$ (orange), $-\gamma$ (blue), and for the 
imaginary parts $-U/\hbar$ (black) and $U/\hbar$ (orange). 
The imaginary parts of both $\lambda_{d,1}$ and $\lambda_{d,{\rm Zeno}}$ are null.
\label{fig:eigenvalues_doublon_U}} 
\end{figure}

One of the remarkable features of this dynamics, clearly noticeable in Fig.~\ref{fig:sx_doublon_U}, is 
the long-time slowing down of the decay rate with increasing dissipative coupling strength. 
Within the two-fermion on two-site model, this slowing down corresponds to the occurrence of
an eigenvalue, $\lambda_{d,\text{Zeno}}$, having a real part proportional 
to $-\frac{24  \gamma J^2}{(\hbar\gamma)^2+4U^2}$ (up to second order in $J$). 
For the case of large dissipative coupling, when $\hbar \gamma \gg U$, the real part of 
this eigenvalue becomes proportional to $-\frac{J^2}{\hbar^2 \gamma}$ and 
corresponds to the previously discussed Zeno effect (see Fig.~\ref{fig:eigenvalues_doublon_U}). 

\begin{figure}[t]
\includegraphics[width=\columnwidth]{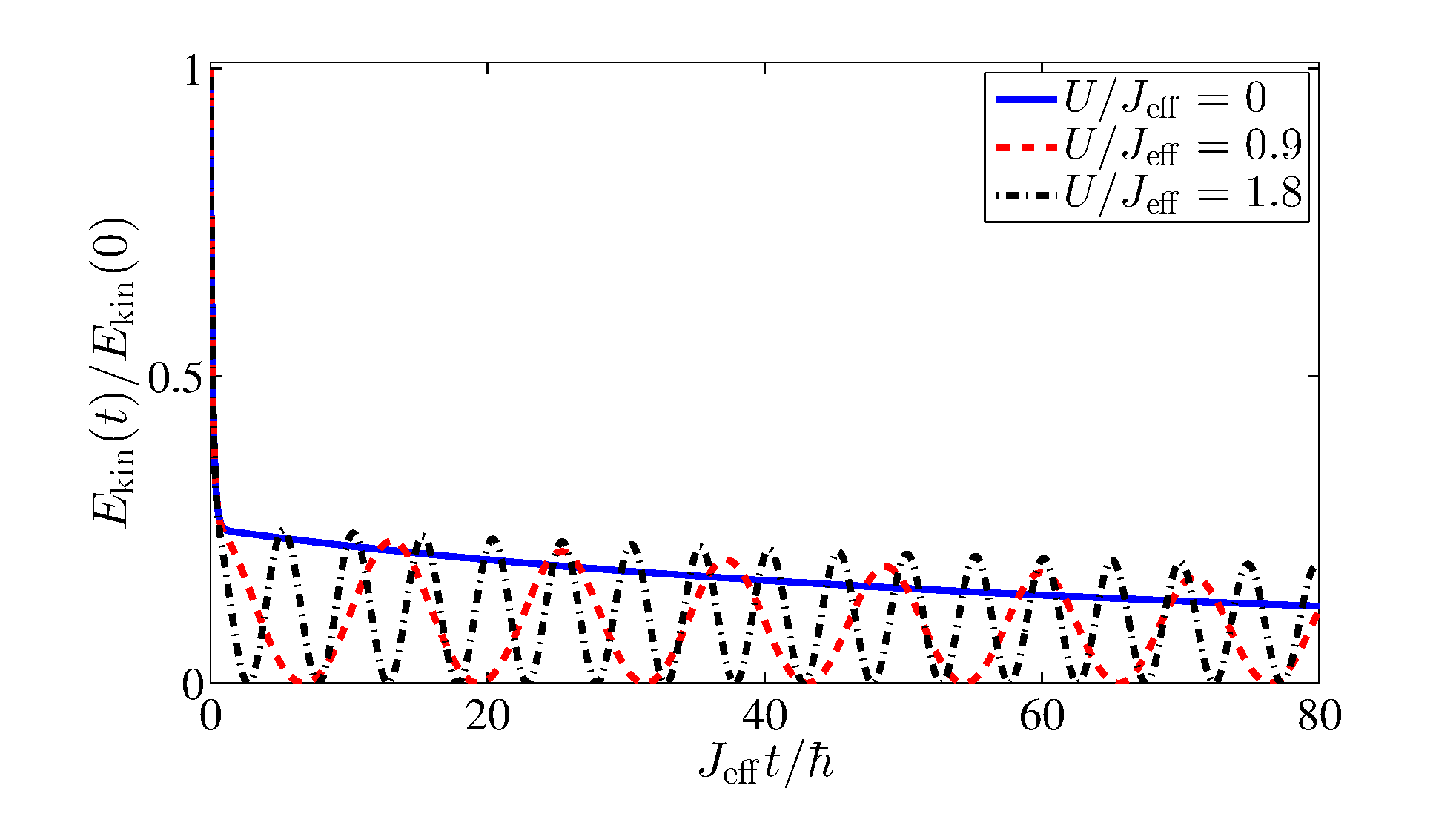}
\caption{(Color online): Time-evolution of the normalized kinetic energy for a dissipative 
coupling $\hbar\gamma/J_{\text{eff}}=10$ to the local double occupancy and different interaction strengths. 
The results are obtained within the slave-spin approach solving Eqs.~(\ref{eq:DOSx}) to (\ref{eq:DOSxSy}).
Here again the evolution begins from the metallic ground state of the mean-field Hamiltonian describing the 
charge sector (Eq.~(\ref{eq:MFH})) with $Z(t=0) = 0,~0.8,~0.098$ for $U/J_{\text{eff}} = 0,~0.8~1.9$ respectively.
\label{fig:sx_doublon_g}} 
\end{figure}

Surprisingly, within the slave-spin approach, the decay rate is found to adopt an algebraic form for $U = 0$. 
For the case of an instantaneous quench to the non-interacting limit, the slave-spin differential equation can 
be solved analytically, and we find the kinetic energy to decay as $\frac{1}{\gamma t}$. 
The numerical simulations at small $\gamma$ provide support for the presence of 
similar algebraic decays even for slightly larger interaction strengths. However, for these cases the identification 
of the exact mathematical form is a much more difficult task. Nevertheless, we can clearly see 
in Fig.~\ref{fig:sx_doublon_g} that the asymptotic 
long-time state at large dissipative couplings or large interaction strengths still possesses a large amount
of kinetic energy, an observation that is in agreement with our predictions obtained in the analytically solvable limits.
An investigation using the adiabatic elimination method could clarify the exact form of the decay 
and the properties of this new steady state; however, due to the high dimensional decoherence free subspace,
which complicates the situation considerably, such a study is beyond the scope of the current work.
 
\begin{figure}[t]
\includegraphics[width=0.7\columnwidth]{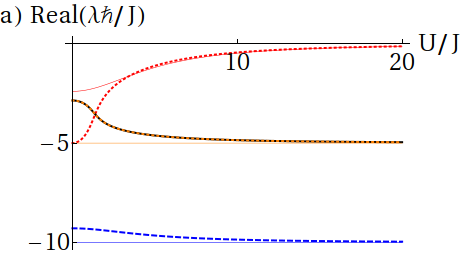}
\includegraphics[width=0.7\columnwidth]{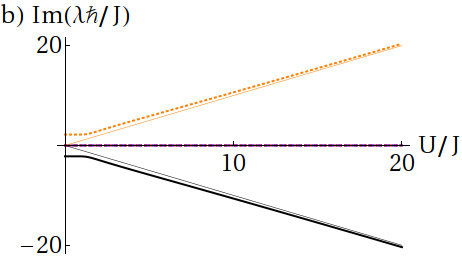}
\caption{(Color online): Real (a) and imaginary (b) parts for $\lambda_{d,1}$ (dashed blue), 
$\lambda_{d,2}$ (black), $\lambda_{d,3}$ (dotted orange) and $\lambda_{n,{\rm Zeno}}$ (dotted red) versus 
$U/J$ for two fermions (one spin up and one spin down) on two sites with dissipative coupling $\hbar\gamma/J=10$ to the
local double occupancy. The solid thin lines correspond to analytical predictions for the real parts $-24\gamma J^2/[(\hbar\gamma)^2+U^2]$ 
(red), $-\gamma/2$ (orange), $\gamma$ (blue), and for the imaginary parts $-U/\hbar$ (black) and $U/\hbar$ (orange). 
The imaginary parts of both $\lambda_{d,1}$ and $\lambda_{d,{\rm Zeno}}$ are null.
\label{fig:eigenvalues_doublon_g}} 
\end{figure}

Finally, the evolution of the kinetic energy within the slave-spin approach is 
characterized by the presence of oscillations at finite interaction strengths.
The frequency of these oscillations is approximately proportional to $U/\hbar$ for very large interaction values as can be seen 
in Fig.~\ref{fig:sx_doublon_g}. However, for small values of $U$, the oscillation frequency decreases with 
increasing values of the dissipative coupling, $\gamma$ (see Fig.~\ref{fig:sx_doublon_U}). 
For the two-fermion on two-site system, this result corresponds to the
saturation, at small $U$, of the imaginary parts of eigenvalues $\lambda_{d,4}$ and $\lambda_{d,Zeno}$ 
as shown in Fig.~\ref{fig:eigenvalues_doublon_g}.

\section{Conclusion}\label{sec:conclusions}
We investigated the dynamics of strongly interacting fermions in a lattice and submitted to dissipative effects.
We studied in detail two kinds of dissipative processes: in the first situation, the environment is effectively coupled to the local fermionic
density whereas, in the second case, the dissipative coupling is to the local double occupancy. As a preamble, we analyzed a system of 
two atoms (one for each spin) on two sites in order to obtain initial insights into the dynamics of these open systems. 
We then developed a novel method to 
study the short-time dynamics of fermionic many-body open systems based on a slave-spin representation of the interacting fermions. 
In parallel, for the case of fermions with a dissipative coupling to the local density, we pursued a radically different approach and explored
their dynamics using adiabatic elimination. This method is designed to work well at long times and complements well the mean-field slave-spin approach. 
We applied these techniques to the two types of dissipative processes, and found that in both cases it is possible to 
highlight the presence of slow decaying states occurring due to the Zeno-effect or to ``interaction impeding'' 
(also referred to as ``interaction Zeno-effect''). We were also able to identify important properties of the steady states. 
In particular, for the coupling to the local density, we found that the  steady state is unique and is the infinite temperature state. 
We also predicted that, for dissipation coupled to the density, the short time dynamics (from the ground state of the fermionic Hamiltonian) 
is characterized by an exponential decay with oscillations whose period dependent on the interaction strength $U$. 
For the local coupling to the double occupancy, we found that, unlike in the previous case, the steady state is not unique and depends 
on the initial condition. The presence of coherence at long times also allows for a more reliable use of the slave-spin method while 
it greatly complicates the use of the adiabatic elimination approach. For this last coupling, we observed an exponentially 
decaying behavior which depends on $U$. The mean-field slave-spin approach also predicts the presence of a power law decay 
when the dissipative system  is quenched to the non-interacting limit. However, this interesting result remains 
to be confirmed using a complementary approach, which is left to future work. To conclude, we believe that the slave-spin 
method developed here, and benchmarked against other reliable approaches, provides a new and exciting framework to study notoriously complex systems: 
interacting lattice fermions coupled to a dissipative environment.   

{\it Note:} During the final stage of this work, a complementary article (Ref.~\onlinecite{SarkarDaley2014}) appeared online. 
The authors of this article derived the Master equation for the fermionic atoms with a coupling to the local density. 
Additionally, an analysis of the two-site and one-dimensional fermionic systems was performed using a combination of the 
stochastic wave function method and of the density-matrix renormalization group.

%%%%%%%%%%%%%%%%%%%%%
% acknowledgments
%%%%%%%%%%%%%%%%%%%%%
\acknowledgments

We thank A.~Georges for his contributions to the early stage of this work. 
We acknowledge BCGS, DFG, CIFAR (Canada), MOE Singapore start-up grant and 
NSERC of Canada for their financial support. 
%%%%%%%%%%%%%%%%%%%%%
%\bibliography{references}

\begin{thebibliography}{35}
\bibitem{footnote1}
Fluorescence scattering is the absorption of 
a (lattice) photon by an atom which is then followed by a spontaneous emission.
\end{thebibliography}

\end{document}